\documentstyle[12pt]{article}
\newcommand{\sss}{\setcounter{equation}{0}}

\newtheorem{theorem}{THEOREM}[section]

\newtheorem{lemma}[theorem]{LEMMA}

\newtheorem{corollary}[theorem]{COROLLARY}

\def\ER{{\bf R}}
\def\beq{\begin{equation}}
\def\ene{\end{equation}}
\def\bull{\begin{flushright} \vrule height 6pt width 6pt depth -.pt
\end{flushright}}

\begin{document}
\baselineskip=14.4pt

\title{$L^p -L^{\acute{p}}$ Estimates for the Schr\"{o}dinger Equation on the
Line and Inverse Scattering for the Nonlinear Schr\"{o}dinger Equation
with
 a Potential \thanks{{\sc ams} classification 35P, 35Q, 35R, 34B and 81U. }}

\author{Ricardo Weder\thanks{Fellow Sistema Nacional de Investigadores.}
\thanks{On leave of absence from IIMAS--UNAM. Apartado Postal 20--726.
M\'exico D.F. 01000. E--Mail: weder@servidor.unam.mx }\\
Instituto de F\'{\i}sica de Rosario,\\
Consejo Nacional de Investigaciones Cient\'{\i}ficas y T\'ecnicas,
\\Bv. 27 de febrero 210 bis, 2000 Rosario, Argentina.\\
E-Mail:weder@ifir.ifir.edu.ar}
\date{}
\maketitle

\begin{center}
\begin{minipage}{5.75in}
\centerline{{\bf Abstract}}\bigskip
In this paper I prove a $L^{p}-L^{\acute{p}}$ estimate for the solutions of
the
one--dimensional Schr\"{o}dinger equation with a potential in $L^{1}_{\gamma}$
where in the {\it generic case} $\gamma > 3/2$ and in the
{\it exceptional case}
 (i.e. when
there is a half--bound state of zero energy) $\gamma > 5/2$. I use this
estimate
 to construct the scattering operator for the nonlinear Schr\"{o}dinger
 equation with a potential. I prove moreover, that the low--energy limit of
 the scattering operator
 uniquely determines the potential and the nonlinearity using a method that
  allows as well for the reconstruction of the potential and of the
   nonlinearity.

\end{minipage}
\end{center}
\newpage

\section{Introduction}\sss
Let us consider the Schr\"{o}dinger equation (LS)
\beq
i\frac{\partial}{\partial t} u(t,x)= H_0 u(t,x),\, u(0,x)=\phi (x)
\label{1.1}
\ene
where $H_0$ is the self--adjoint realization of $-\Delta$ in
$L^2\left(\ER^n \right), n \geq 1,$

\beq
H_0 := -\sum^{n}_{j=1}\frac{\partial^2}{\partial x^2_j }.
\label{1.2}
\ene
 The domain of $H_0, D(H_0 ),$ is the Sobolev space $W_2$. The solution to
(\ref{1.1}) is given by $e^{-itH_0} \phi$, where the strongly continuous
unitary group $e^{-itH_0}$ is defined by the functional calculus of
 self--adjoint operators. The kernel of $e^{-itH_0}$ is given by ( see
 Example 3 in page 59 of \cite{6} ) $(4 \pi it )^{-n/2} e^{i|x-y|^2 /4 t}$.
 From this explicit expression for the kernel it follows that the restriction
 of $e^{-it H_0}$ to $L^2 (\ER^n )\cap L^p (\ER^n )$ extends to a bounded
 operator from $L^p (\ER^n ) $ into $L^{\acute{p}} (\ER^n )$ such that
 \beq
 \left\| e^{-it H_0}\right\|_{{\cal B}\left( L^p (\ER^n ), L^{\acute{p}}(\ER^n )
 \right)} \leq \frac{C}{t^{n(\frac{1}{p}-\frac{1}{2})}},\, t > 0,
 \label{1.3}
 \ene
 for some constant $C$, $1 \leq p \leq 2,$ and $\frac{1}{p}
 +\frac{1}{\acute{p}}=1,$ and where for any pair of Banach spaces $X,Y$ we
 denote by ${\cal B}(X, Y )$ the Banach space of all bounded linear operators
 from $X$ into $Y$. In the case when $X=Y$ we use the notation ${\cal B}(X)$.
 Estimate (\ref{1.3}) expresses the dispersive nature of the solutions to
 (\ref{1.1}) and it is a fundamental tool in the study of the nonlinear
 Schr\"{o}dinger equation:

 \beq
 i\frac{\partial}{\partial t}u= H_0 u + f(u)
 \label{1.4}
 \ene
 since it allows to control the nonlinear behaviour of the solutions to
 (\ref{1.4}), that is produced by $f(u)$, in terms of the dispersion that is
 produced by the linear term $H_0 u$. See for example \cite{6},  \cite{20},
 \cite{21}, \cite{22}, \cite{15},
 \cite{16}, \cite{18}, \cite{19},
  \cite{33},
 \cite{27} and \cite{32}.

 In the case of a linear Schr\"{o}dinger equation with a potential (LSP):
 \beq
 i\frac{\partial}{\partial t} u(t,x)=(H_0 +V)u(t,x), \, u(0,x)=\phi,
 \label{1.5}
 \ene
 where $V$ is a real--valued function defined on $\ER^n$ such that the operator
 $H:= H_0 +V$ is self--adjoint on $D(H_0 )$, Journ\'e, Soffer and Sogge
 \cite{10}
 proved
 that for $n \geq 3$

 \beq
\left\|e^{-it H_0} P_c \right\|_{{\cal B}\left(L^p (\ER^n ), L^{\acute{p}}
(\ER^n )\right)} \leq \frac{C}{t^{n(\frac{1}{p}-\frac{1}{2})}},
 \label{1.6}
 \ene
 for $1\leq p \leq 2,\,\frac{1}{p}+\frac{1}{\acute{p}}=1$ and where $P_c$ is
 the orthogonal projector onto the continuous subspace of $H$. Note that
 (\ref{1.6}) can not hold for the pure point subspace of $H$. Estimate
 (\ref{1.6}) is the natural extension of (\ref{1.3}) to the case with a
 potential. Besides conditions on the regularity and the decay of $V$ (see
 equation (1.6) of \cite{10}) Journ\'e, Soffer and Sogge require that zero is
 neither a bound state nor a half--bound state for $H$. The proof given by
 \cite{10} consists of a high--energy estimate that is always true and of a
 low--energy estimate where the condition that zero is neither a bound state nor a
 half--bound state was used. The low--energy estimate of \cite{10} was obtained
 by studying the behaviour near zero of the spectral family of $H$. For this
 purpose Journ\'e, Soffer and Sogge \cite{10} used the estimates on the
 behaviour near
 zero of the resolvent of $H$ obtained by Jensen and Kato \cite{34}, \cite{30}
 and \cite{31} for $n\geq 3$. It is actually here that the restriction $n \geq
 3 $ appears in the result of \cite{10}. One way to understand the
 reasons for the restriction to $n \geq 3$ is to look to the kernel of the
 free resolvent, $\left(H_0 -z \right)^{-1}$. For $n=3$ this kernel is given by

 \beq
 \frac{1}{4\pi} \frac{e^{i\sqrt{z}\, |x-y|}}{|x-y|}.
 \label{1.7}
 \ene
 Note that (\ref{1.7}) behaves nicely as $z\rightarrow 0$. In the case $n \geq
 4$ the kernel of the free resolvent has also a nice behaviour as
 $z \rightarrow 0$. This fact is the starting point of the analysis of Jensen
 and Kato in \cite{34}, \cite{30} and \cite{31}, who use perturbation theory to
 estimate the behaviour near zero of the resolvent of $H$. In the case $n=1$
 the kernel of $\left( H_0 -z \right)^{-1}$ is given by (see Theorem 9.5.2 in
 page 160 of \cite{13})
 \beq
 \frac{i}{2\sqrt{z} } e^{i\sqrt{z}|x-y|}.
 \label{1.8}
 \ene
 The kernel ({\ref{1.8}) is singular as $z\rightarrow 0$ and an approach as in
 \cite{10}, \cite{34}, \cite{30} and \cite{31} does not appears to be
 convenient. We take in Section 2 below  a different point of view. We
 base our analysis of the low--energy behaviour of the spectral family of $H$
 on the generalized Fourier maps that are constructed from the scattering
 solutions $\Psi_+ (x,k), x,k \in \ER$. The crucial issue here is that for
 $n=1$ the construction of the scattering solutions can be reduced to the
 solution of Volterra integral equations. More precisely, the scattering
 solution is given in terms of the Jost solutions, $f_j (x,k), j=1,2,$ as
 follows:
\beq
\Psi_+ (x,k)=\cases{\frac{T(k)}{\sqrt{2\pi}} f_1 (x,k), & $k \geq 0,$\cr\cr
\frac{T(-k)}{\sqrt{2\pi}} f_2 (x,-k),& $k\leq 0,$}
\label{1.9}
\ene
where $T(k)$ is the transmission coefficient. The $f_j$ are solutions to
Volterra integral equations that are obtained by iteration as uniformly
convergent series. See \cite{25}, \cite{26}, \cite{1} and \cite{28}. This fact
allows for a detailed analysis of the low--energy behaviour of the spectral
family of $H$ that coupled with a high--energy estimate allows us to prove in
Section 2 an estimate like (\ref{1.6}) in the case $n=1$.

Since in what
follows we only consider the case $n=1$ we denote below by $L^p ,1 \leq p \leq
\infty,$ the space $L^p \left(\ER^1 \right)$. For any $s\in \ER$ let us denote
by $L^1_s$ the space of all complex--valued measurable functions, $\phi $,
defined on $\ER$ such that
\beq
\|\phi\|_{L^1_s}:=\int_{\ER} |\phi (x)| (1+|x|)^s dx < \infty.
\label{1.10}
\ene
$L^1_s$ is a Banach space with the norm (\ref{1.10}).
Below we always assume that $V\in L^1_1$. It follows from the existence of the
Jost solutions and since the eigenvalues of $-\frac{d^2}{d x^2}+V(x)$ are
simple (see \cite{1}) that the differential expression
$\tau:= -\frac{d^2}{d x^2}+V(x)$ is in the limit point case at $\pm \infty$.
Then by the Weyl criterion (see \cite{12}) $\tau$ is essentially
self--adjoint on the domain
\beq
D(\tau ):=\left\{ \phi \in L^2_C : \phi \,\hbox{and}\, \acute{\phi}
\hbox{ are absolutely
continuous and $\tau \phi \in L^2$ }\right\},
\label{1.11}
\ene
where we denote by $\acute{\phi}(x)= \frac{d}{d x}\phi(x)$
and by $L^2_C$ the set of all $\phi \in L^2$ that have compact support.
We denote by $H$ the unique self--adjoint realization of $\tau $. It is known
that the absolutely continuous spectrum of $H$ is given by $\sigma_{ac}(H)=
[0, \infty ),$ that $H$ has no singular continuous spectrum, that $H$ has no
eigenvalues that are positive or equal to zero and that $H$ has a finite
number, $N$, of negative eigenvalues that are simple and that we denote by
$-\beta^2_N < \beta^2_{N-1} < \cdots < -\beta^2_1 < 0$. Let $F$ denotes the
Fourier
transform as a unitary operator on $L^2$
\beq
F\phi (k)=\frac{1}{\sqrt{2\pi}} \int_{-\infty}^{\infty}
e^{-ikx}\phi (x) dx.
\label{1.12}
\ene
We will also use the notation $\hat{\phi }(k):= F\phi (k)$.
For any
$\alpha \in  \ER$ let us denote by $W_{\alpha}$ the Sobolev space consisting
of the completion of the Schwartz class in the norm
\beq
\left\|\phi \right\|_{\alpha}:=\left\|(1+k^2)^{\alpha /2}
\hat{\phi }(k)\right\|_{L^2}
.
\label{1.13}
\ene
We denote by $h$ the following quadratic form
\beq
h(\phi ,\psi):= (\acute{\phi}, \acute{\psi})_{L^2} + (V\phi , \psi )_{L^2},
\label{1.14}
\ene
with domain $D(h)= W_1 $. Since $V\in L^1_1 \subset L^1_0 \equiv L^1$ it
follows from Theorem 8.42 in page 147 of \cite{13} and from
the remarks above Theorem 9.14.1 in page 183 of \cite{13} that $h$ is
closed and
bounded from below and that the associated operator, $H_h$, is self--adjoint
with domain, $D(H_h )\subset W_1$. Since $D(\tau ) \subset W_1 $ it follows
that $H_h$ is a self--adjoint extension of $\tau$ and as $\tau$ is essentially
self--adjoint we have that $H=H_h$ and then
$
D(|H|)=W_1
$.
For $u, v$ any pair of solutions to the stationary Schr\"{o}dinger equation:
\beq
-\frac{d^2}{d x^2}u + V u=k^2 u,\,\, k\in \ER,
\label{1.16}
\ene
let $[u,v]$ denotes the Wronskian of $u$ and $v$:
\beq
[u,v]:= \acute{u} v -u \acute{v}.
\label{1.17}
\ene
A potential $V$ is said to be {\it generic } if the Jost solutions at zero
energy satisfy \linebreak$[f_1 (x,0) ,f_2 (x,0)]\neq 0$ and $V$ is said
to be {\it
exceptional }if $[f_1 (x,0), f_2 (x,0)]=0$. If the potential $V$ is exceptional
there is a bounded solution (a half--bound state ) to the equation
(\ref{1.16}) with $k=0$. See \cite{2} for these definitions and a discussion of
related issues. Let $P_c$ denotes the projector onto the continuous subspace
of $H$. Note that $P_c =I-P_p$, where $P_p$ is the projector onto the
finite dimensional subspace of $L^2$ generated by the eigenvectors
corresponding to the $N$ eigenvalues of $H$.

Our mail result is the following theorem that we prove in Section 2.
\begin{theorem}(The $L^1 - L^{\infty}$ estimate ).
Suppose that $V\in L^1_{\gamma}$ where in the {\rm generic case} $\gamma > 3/2$
and in the {\rm exceptional case} $\gamma > 5/2$. Then for some constant $C$
\beq
\left\|e^{-it H}P_c \right\|_{{\cal B}\left(L^1 ,L^{\infty}\right)}
\leq \frac{C}{\sqrt{t}}, \, t > 0.
\label{1.18}
\ene
\end{theorem}
\begin{corollary} (The $ L^p -L^{\acute{p}}$ estimate).
 Suppose that the conditions of Theorem 1.1 are satisfied. Then for
$1 \leq p \leq 2$ and $\frac{1}{p}+\frac{1}{\acute{p}}=1$
\beq
\left\|e^{-itH} P_c \right\|_{{\cal B}\left( L^p ,L^{\acute{p}}\right)}
\leq \frac{C}{t^{(\frac{1}{p}-\frac{1}{2})}}, \, t > 0.
\label{1.19}
\ene
\end{corollary}
\begin{corollary} (The espace--time estimate). Suppose that the conditions
of Theorem 1.1 are satisfied. Then

\noindent(a)
\beq
e^{-itH}P_c \in {\cal B}\left(L^2 ,L^6 (\ER \times \ER )\right).
\label{1.20}
\ene

\noindent (b) If moreover, $H$ has no negative eigenvalues and
\beq
i \frac{\partial}{\partial t}u(t,x) = H u(t,x) +g(t,x), \, u(0,x)=f(x),
\label{1.21}
\ene
then
\beq
\left\|u(t,x)\right\|_{L^6 (\ER \times \ER )} \leq C \left[ \|f\|_{L^2}
+\|g\|_{L^{6/5} (\ER \times \ER )}\right].
\label{1.22}
\ene
\end{corollary}
In the case $V=0$ and $n \geq 1$ Theorem 1.1 and Corollaries 1.2 and 1.3
were proven by Strichartz in \cite{11}. They were proven in \cite{10}
for $n \geq 3$ and $V$ satisfying appropriate conditions on regularity and
decay (see \cite{10}, equation (1.6)). In \cite{10} it was assumed moreover,
that zero is neither a bound state nor a half--bound state. Note that we do not
have to assume that zero is not a half--bound state for Theorem 1.1 and
Corollaries 1.2 and 1.3 to hold. In our case it is enough to require that
$V$ has a
slightly  faster decay at infinity when there is a half--bound state at zero.

Theorem 1.1 and Corollaries 1.2 and 1.3 open the way to the study of the
 scattering theory for the nonlinear Schr\"{o}dinger equation
with a potential (NLSP):
\beq
i\frac{\partial}{\partial t}u = H u + f(|u|) \frac{u}{|u|}.
\label{1.23}
\ene
As a first application we study in this paper the low--energy scattering for
the NLSP and we prove that the low--energy limit of the scattering
operator uniquely determines the
potential and the nonlinearity. For this purpose we proceed as in \cite{23}
were the case $n \geq 3 $ was considered. Let us assume that $H$ has no
negative eigenvalues. Then $H >0$ and since $D(\sqrt{H})=W_1$ the operators
$\sqrt{H+1}\, (-\Delta +1)^{-1/2}$ and $\sqrt{-\Delta +1}\, (H+1)^{-1/2}$ are
bounded in $L^2$. It follows that the norm associated to the following
scalar product
\beq
 (\phi ,\psi )_{X}:=\left( \sqrt{H+1} \,\phi , \sqrt{H+1} \,\psi \right)_{L^2},
 \label{1.24}
 \ene
 is equivalent to the norm of $W_1$. We denote by $X$ the Sobolev space $W_1$
 endowed with the scalar product (\ref{1.24}). The space $X$ is a Hilbert
 space. Clearly, $e^{-itH}$ is a strongly continuous group of unitary operators
 on $X$. For any $\delta > 0$ we denote:

 \beq
 X(\delta ):=\left\{ \phi \in X : \|\phi \|_{X} < \delta \right\}.
 \label{1.25}
 \ene
 Let us denote  $X_3 := L^{p+1}$ and $r=(p-1)/(1-d)$ with
 $d:=\frac{1}{2} (p-1)/
 (p+1)$ and $5 \leq p < \infty$. In what follows for functions $u(t,x)$ defined
 on $\ER \times \ER$ we write $u(t)$ for $u(t,\cdot )$.

 \begin{theorem} (Low--energy scattering). Suppose that $V \in L^1_{\gamma}$
 where in the {\rm generic case} $ \gamma > 3/2$ and in the
 {\rm exceptional case} $\gamma
 > 5/2$ and that $H$ has no negative eigenvalues. Assume moreover, that the
  function $f$ in (\ref{1.23}) is defined on $\ER$, that it is real--valued
  and $C^1$. Furthermore, $f(0)=0$ and
  \beq
  \left| \frac{d}{d \mu}f(\mu )\right| \leq C |\mu |^{p-1},
  \label{1.26}
  \ene
  for some  $ 5 \leq p < \infty $. Then there is a $\delta > 0$ such that for
  every $\phi_- \in X(\delta )$ there is a unique solution to the NLSP,
  $u(t,x)$, such that $u\in C(\ER , X)\cap L^r (\ER , X_3 )$ and
  \beq
  \lim_{t \rightarrow - \infty} \left\| u(t) -e^{-itH} \phi_- \right\|_X =0.
  \label{1.27}
  \ene
  Moreover, there exists a unique $\phi_+ \in X$ such that
  \beq
  \lim_{t \rightarrow \infty}\left\|u(t)- e^{-itH}\phi_+ \right\|_X=0.
  \label{1.28}
  \ene
  For all $t\in \ER$
  \beq
  \frac{1}{2} \|u(t)\|^2_X +\int_{\ER} F(|u(t)|) dx =\frac{1}{2}
  \|\phi_- \|^2_X
  =\frac{1}{2}\|\phi_+\|^2_X,
  \label{1.29}
  \ene
  where $F$ is the primitive of $f$ such that $F(0)=0$. In addition the
  nonlinear scattering operator $S_V : \phi_- \rightarrow \phi_+$ is a
  homeomorphism from $X(\delta )$ onto $X(\delta )$.
  \end{theorem}
 Theorem 1.4 is proven in Section 3 using Theorem 1.1, Corollaries 1.2 and 1.3
 and the abstract low--energy scattering theory of Strauss \cite{15},
 \cite{16}.
The scattering operator $S_V$ compares solutions of the NLSP (\ref{1.23}) with
 solutions to the LSP (\ref{1.5}). To reconstruct $V$ we consider below the
 scattering operator, $S$, that compares solutions to the NLSP with solutions
 to the LS (\ref{1.1}). For this purpose let us  consider the wave operators
 \beq
 W_{\pm}:= \hbox{s}-\lim_{t \rightarrow \pm \infty} e^{itH}\, e^{-itH_0}.
 \label{1.30}
 \ene
 The $W_{\pm}$ are unitary on $L^2$ (note that $H$ has no eigenvalues). The
 existence of the strong limits in (\ref{1.30}) is well known (see Theorem
 9.14.1 in page 183 of\cite{13}). Moreover, by the intertwining relations,
 $\sqrt{H} W_{\pm}= W_{\pm}\sqrt{H_0}$ and as $D(\sqrt{H})=W_1$, we have that
 $W_{\pm}$ and $W^{\ast}_{\pm}$ belong to ${\cal B}(W_1)$ and for
 $0 < \delta_1
 < \delta$
  they send $X(\delta_1 )$ into $X(\delta )$ if $\delta_1$ is small
 enough. Let us define:
 \beq
 S:= W^{\ast}_+ S_V W_- .
 \label{1.31}
 \ene
 Take $\delta_1$ so small that $W_- X(\delta_1 )\subset X(\delta )$ with
 $\delta$ as in Theorem 1.4 and then $\delta_2$ so large that $W^{\ast}_+
 X(\delta ) \subset X( \delta_2 )$. Then $S$ sends $X(\delta_1 )$ into
 $X(\delta_2 )$. Moreover, for any $\psi_- \in X(\delta_1 )$ let us take in
 Theorem 1.4
 $\phi_- \equiv W_- \psi_-$  and let $u(t,x)$ and $\phi_+$ be as in
 Theorem 1.4.
 Let us denote $\psi_+ :=S \psi_- = W^{\ast}_+ \phi_+$. Then by Theorem 1.4 and
 (\ref{1.30})
 \beq
 \lim_{t \rightarrow \pm \infty}\left\|u(t,x)-e^{-itH_0}\psi_{\pm}
 \right\|_{L^2}=0.
 \label{1.32}
 \ene
 That is to say, $S$ sends the initial data at $t=0,\psi_-  ,$ of the
 incoming solution to LS to the initial data at $t=0, \psi_+ ,$ of the
 outgoing solution to LS. Let us denote by $S_L$ the linear scattering operator
 corresponding to the LS and the LSP:
 \beq
 S_L := W^{\ast}_+ \,W_-.
 \label{1.33}
 \ene
 In Theorem 1.5 below, $S_L$ is reconstructed from the low--energy limit
 of $S$.
 \begin{theorem}
 Suppose that the assumptions of Theorem 1.4 are satisfied. Then for every
 $\phi ,\psi \in X$
 \beq
 \lim_{\epsilon \downarrow 0}\left(S \epsilon \phi , \psi \right)_{L^2}=
 \left( S_L \phi ,\psi \right)_{L^2}.
 \label{1.34}
 \ene
 \end{theorem}
 Since, as is well known, from $S_L$ we can uniquely reconstruct $V$ we
 obtain the following Corollary.
 \begin{corollary}
 Suppose that the assumptions of Theorem 1.5 are satisfied. Then the scattering
 operator, S, uniquely determines the potential $V$.
 \end{corollary}
 In the case where $f(u)= \lambda |u|^p$, we can also uniquely reconstruct the
 coupling  constant $\lambda$.
 \begin{corollary}
Suppose that the assumptions of Theorem 1.4 are satisfied and that moreover,
$f(u)=\lambda |u|^p $, for some constant $\lambda$. Then the scattering
operator, S, uniquely determines the potential $V$ and the coupling constant
$\lambda$. Furthermore, for all $0\neq \phi \in X \cap L^{1+\frac{1}{p}}$:
\beq
\lambda =\lim_{\epsilon \downarrow 0} \frac{1}{\epsilon^p} \frac{\left(
(S_V -I)\epsilon \phi ,\phi \right)_{L^2}}{\int^{\infty}_{-\infty}
\left\|e^{-itH} \phi \right \|^{1+p}_{L^{1+p}}}.
\label{1.35}
\ene
\end{corollary}
Remark that by Sobolev's imbedding theorem \cite{14}, $X \subset L^{1+p}$. Then
by (\ref{1.19})
\beq
0 < \int^{\infty}_{-\infty} \left\| e^{-itH} \phi \right\|^{1+p}_{L^{1+p}} dt
< \infty.
\label{1.36}
\ene
 Theorem 1.5 and Corollaries 1.6 and 1.7 are proven
 as in \cite{23} (see Section 3).

 We use below the letter $C$ to denote any
 positive constant whose particular value is not relevant.

\section{The $L^p -L^{\acute{p}}$ Estimate}\sss
We assume that $V\in L^1_1$. For any complex number, $k$,
 we denote by $\Re k$ and $\Im k$, respectively, the real and the imaginary
 parts of $k$. The Jost solutions $f_j (x,k), j=1,2,$
are solutions
to the stationary Schr\"{o}dinger equation
\beq
-\frac{d^2}{d x^2} f_j (x,k) +V(x) f_j (x,k)= k^2 f_j (x,k)
\label{2.1}
\ene
 were $\Im k \geq 0$. To construct the Jost solution we define $ m_1 (x,k):=
 e^{-ikx} f_1 (x,k)$ and $m_2 (x,k) :=e^{ikx} f_2 (x,k)$. They are,
 respectively, solutions of the following equations:

 \beq
 \frac{d^2}{d x^2} m_1 (x,k) +2ik \frac{d}{d x} m_1 (x,k)= V(x) m_1 (x,k),
 \label{2.2}
 \ene
 \beq
 \frac{d^2}{d x^2} m_2 (x,k) -2ik \frac{d}{d x} m_2 (x,k)= V(x) m_2 (x,k).
 \label{2.3}
 \ene
The $m_j (x,k), j=1,2,$ are the unique solutions of the Volterra integral
equations
\beq
m_1(x,k)= 1 + \int_x^{\infty} D_k (y-x) V(y) m_1 (y,k) dy,
\label{2.4}
 \ene
 \beq
 m_2 (x,k)= 1+\int^x_{-\infty} D_k (x-y) V(y) m_2 (y,k) dy,
 \label{2.5}
 \ene
 where
 \beq
 D_k (x):=\int_0^x e^{2iky}dy =
 \cases{ \frac{1}{2ik} (e^{2ikx}-1),& $k\neq 0,$ \cr\cr
 x,& $k=0.$}
 \label{2.6}
 \ene
 Note that $f_1 (x,k) \sim e^{ikx}$ as $ x \rightarrow \infty$ and that
 $f_2 (x,k)\sim e^{-ikx}$ as $ x\rightarrow -\infty$. A detailed study of the
 properties of the $m_j (x,k),j=1,2,$ was  carried over in \cite{1}. Here we
  state a number of results from \cite{1} that we need. In what follows we
  denote by $C$ any positive constant whose specific value is not relevant to
  us and by $\dot{g}(x,k):= \frac{\partial}{\partial k}g(x,k)$.
  For each fixed $x \in \ER$ the $m_j (x,k)$ are analytic in $k$ for $\Im k > 0$
  and continuous in $\Im k \geq 0$ and
  \beq
  |m_1 (x,k) -1 | \leq C\,\, \frac{1+\hbox{max} (-x ,0)}{1+|k|},
  \label{2.7}
  \ene
  \beq
  |m_2 (x,k) -1| \leq C\,\, \frac{1+\hbox{max}(x,0)}{1+|k|}.
  \label{2.8}
  \ene
  Moreover, $\dot{m}_j (x,k),j=1,2,$
  exits for $\Im k\geq 0, k\neq 0$, $ k\dot{m}_j (x,k)$ is continuous in $k$ for
   each
  fixed $x\in \ER $ and for each fixed $x_0\in \ER$ there is a constant
  $C_{x_0}$  such that
  \beq
  |\dot{m}_1 (x,k)|\leq C_{x_0} \,\,\frac{1}{|k|}, x \geq x_0,
  \label{2.10}
  \ene
  \beq
  |\dot{m}_2 (x,k)| \leq C_{x_0} \,\,\frac{1}{|k|}, x \leq x_0 .
  \label{2.11}
  \ene
  In the Lemma below we slighly improve the estimates (\ref{2.10}) and
  (\ref{2.11}) under the assumption that $V \in L^1_{\gamma}$ for
  $1 < \gamma \leq 2$.
  \begin{lemma}
  Suppose that $V \in L^1_{\gamma}$ for some $1 \leq \gamma \leq 2$. Then for
  each $x_0 \in \ER$ there is a constant $C_{x_0}$ such that
  \beq
|\dot{m}_1 (x,k)| \leq C_{x_0}\, \frac{|k|^{\gamma}}{|k|^2 (1+|k|)^{\gamma-1}}
, x \geq x_0,
\label{2.12}
\ene
\beq
|\dot{m}_2 (x,k)| \leq C_{x_0} \,\frac{|k|^{\gamma}}{|k|^2 (1+|k|)^{\gamma -1}}
, x \leq x_0 .
\label{2.13}
\ene
\end{lemma}
\noindent{\it Proof :}\, We give the proof in the case of $\dot{m}_1 (x,k)$.
The case of $\dot{m}_2 (x,k)$ follows similarly. It follows from (\ref{2.6})
that for $k\neq 0$
\beq
\left|\dot{D}_k (x)\right|=\left|\frac{1}{k}\int_0^{x} y \left(
\frac{\partial}{\partial y}e^{2iky}\right) \, dy \right|\leq 2 \frac{|x|}{|k|},
\label{2.14}
\ene
and that
\beq
\left|\dot{D}_k (x)\right| \leq |x|^2.
\label{2.15}
\ene
By (\ref{2.14}) and (\ref{2.15}) for any $1\leq \gamma \leq 2$
\beq
|\dot{D}_k (x)|\leq \frac{2^{2-\gamma} |x|^{\gamma}}{|k|^{2-\gamma}}.
\label{2.15b}
\ene
Since (\ref{2.4}) is a Volterra integral equation, $m_1 (x,k)$ is obtained by
iteration \cite{1}:
\beq
m_1 (x,k)=\lim_{n \rightarrow \infty} m_{1,n}(x,k),
\label{2.16}
\ene
where $m_{1,0}(x,k)=1$ and for $n=1,2,\cdots$
\beq
m_{1,n}(x,k)=1+\sum_{l=1}^{n}g_l (x,k),
\label{2.17}
\ene
where
\beq
g_l (x,k)=\int_{x\leq x_1 \leq x_2 \leq \cdots \leq x_l}D_k (x_1 -x)
D_k (x_2 -x_1 ) \cdots D_k (x_l -x_{l-1}) V(x_1 )\cdots V(x_l )dx_1
\cdots dx_l.
\label{2.18}
\ene
Moreover, the $m_{1,n}$ satisfy the following equation for $n=0,1,\cdots$
\beq
m_{1,n+1}(x,k)=1+\int_x^{\infty}D_k (y-x) V(y) m_n (y,k) dy.
\label{2.19}
\ene
Then,
\beq
\dot{m}_{1,n+1}(x,k)=\int_x^{\infty}\dot{D}_k (y-x) V(y) m_n (y,k) dy
+\int_x^{\infty}D_k (y-x)V(y) \dot{m}_n (y,k) dy.
\label{2.20}
\ene
Furthermore, since by (\ref{2.6})
\beq
|D_k (x)|\leq |x|,
\label{2.21}
\ene
 it follows from (\ref{2.18}) that
 \beq
 |g_l (x,k)| \leq \frac{1}{l!} \left(\int_x^{\infty} (y-x) V(y) dy \right)^l,
 \label{2.22}
 \ene
 and then by (\ref{2.17}) for $x \geq x_0$

 $$
 |m_{1,n}(x,k)|\leq 1+ \sum_{l=1}^n \frac{1}{l!} \left(\int_x^{\infty} (y-x)
 |V(y)| dy\right)^l
$$
 \beq
 \leq  e^{(\int_x^{\infty} (|x_0 |+|y|) |V(y)| dy)}, \,
 x \geq x_0.
 \label{2.23}
 \ene
 We can now estimate the first integral in the right--hand side of
 (\ref{2.20}) as follows

$$
\left| \int_x^{\infty}\dot{D}_k (y-x) V(y) m_n (y,k) dy \right|\leq
$$
\beq
\frac{2^{2-\gamma}}{|k|^{2-\gamma}}\int_x^{\infty} |y-x|^{\gamma}\, |V(y)|
e^{ \int_x^{\infty}(|x_0 |+|y|) |V(y)| dy}\leq C \,\frac{1}{|k|^{2-\gamma}},
x \geq x_0,
\label{2.24}
\ene
 where we used (\ref{2.15b}). Then using again (\ref{2.20}) and (\ref{2.21})
 we obtain that
 \beq
 |\dot{m}_{1,n+1}(x,k)|\leq \, \frac{C}{|k|^{2-\gamma}} + \int_x^{\infty}
 |y-x| |V(y)| |\dot{m}_n (y,k)| dy.
 \label{2.25}
 \ene
 Since $m_0 (y,k) \equiv 1$ it follows from (\ref{2.25}) with $n=0$ that
 \beq
 |\dot{m}_{1,1}(x,k)|\leq \frac{C}{|k|^{2-\gamma}}.
 \label{2.26}
 \ene
 Then by (\ref{2.25}) we have that
 \beq
 |\dot{m}_{1,2}(x,k)|\leq\, \frac{C}{|k|^{2-\gamma}} (1+ q(x)), x \geq x_0,
 \label{2.27}
 \ene
 where
 \beq
 q(x):=  \int_{x}^{\infty} (|x|_0 +|y|) |V(y)| dy,
 \label{2.28}
 \ene
 and then, iterating (\ref{2.25}) $n-1$ more times we prove that
 \beq
 |\dot{m}_{1,n+1}(x,k)|\leq \frac{C}{|k|^{2-\gamma}}\sum_{l=0}^n
 \frac{(q(x))^l}{l!}.
 \label{2.29}
 \ene
 Taking the limit as $n \rightarrow \infty$ in (\ref{2.29}) we prove that
 \beq
 |\dot{m}_1 (x,k)| \leq \frac{C}{|k|^{2-\gamma}} e^{q(x)}, x \geq x_0.
 \label{2.30}
 \ene
 Since $V\in L^1_{\gamma} \subset L^1_1 $, we can take $\gamma =1$ in
 (\ref{2.30}) and then

 \beq
 |\dot{m}_1 (x,k)|\leq \frac{C}{|k|} \, e^{q(x)},\, x \geq x_0.
 \label{2.31}
 \ene
 Equation (\ref{2.12}) follows from  (\ref{2.30}) and (\ref{2.31}).

 \begin{corollary}
 Suppose that $V \in L^1_{\gamma}$, for some $1\leq \gamma \leq 2$. Then for each
 $x_0 \in \ER$ there is a constant $C_{x_0}$ such that for all $\Im k \geq 0$
 \beq
 \left|\dot{\acute{m}}_1 (x,k)\right|\leq C_{x_0} \left[ 1+\frac{|k|^{\gamma}}
 {|k|^2 (1+|k|)^{\gamma -1}}\right], x \geq x_0 ,
 \label{2.32}
 \ene
 \beq
 \left|\dot{\acute{m}}_2 (x,k)\right| \leq C_{x_0} \left[1+
 \frac{|k|^{\gamma}}{|k|^2 (1+|k|)^{\gamma -1}}\right], \, x \leq x_0.
 \label{2.33}
 \ene
 \end{corollary}

 \noindent {\it Proof :}\, We prove (\ref{2.32}). The proof of (\ref{2.33})
 is similar. By (\ref{2.4}) and (\ref{2.6})
 \beq
 \acute{m}_1 (x,k)= -\int_{x}^{\infty} e^{2ik(y-x)} V(y) m_1 (y,k)  dy,
 \label{2.34}
 \ene
 and then
 \beq
 \dot{\acute{m}}_1 (x,k) =-\int_x^{\infty} \left[2i e^{2ik(y-x)} (y-x) V(y)
 m_1 (y,k)  + e^{2ik(y-x)} V(y) \dot{m}_1 (y,k)\right] dy.
 \label{2.35}
 \ene
 It follows from (\ref{2.7}), (\ref{2.12}) and (\ref{2.35}) that
 \beq
 \left|\dot{\acute{m}}_1 (x,k)\right| \leq C_{x_0} \left[ 1+
 \frac{|k|^{\gamma}}{|k|^2 (1+|k|^{\gamma -1})}\right], x \geq x_0.
 \label{2.36}
 \ene

 \begin{lemma}
 Suppose that $V\in L^1_{\gamma}$, for some $2\leq \gamma \leq 3$. Then for
 every
 $x_0 \in \ER $ there is a constant $C_{x_0}$ such that
 \beq
 \left|\dot{m}_1 (x,k)-\dot{m}_1 (x,0)\right|\leq C_{x_0} |k|^{\gamma -2},
 x \geq x_0,
 \label{2.37}
 \ene
\beq
\left|\dot{m}_2 (x,k) -\dot{m}_2 (x,0)\right|\leq C_{x_0} |k|^{\gamma -2},
x \leq x_0.
\label{2.38}
\ene
\end{lemma}
\noindent{\it Proof:}\, It follows from the definition of $D_k (x)$ in
(\ref{2.6}) that
\beq
\left|\dot{D}_k (x) -\dot{D}_0 (x)\right|\leq \frac{4}{3} |k| |x|^3,
\label{2.39}
\ene
and that
\beq
\left|\dot{D}_k (x) -\dot{D}_0 (x)\right| \leq 2 |x|^2.
\label{2.40}
\ene
Then for any $2\leq \gamma \leq 3$ there is a constant, $C_{\gamma}$, such that
\beq
\left|\dot{D}_k (x)-\dot{D}_0 (x)\right|\leq  C_{\gamma} |k|^{\gamma-2}
|x|^{\gamma}.
\label{2.41}
\ene
We obtain from (\ref{2.20}) that
$$
\dot{m}_{1,n+1}(x,k)- \dot{m}_{1,n+1}(x,0)=\int_x^{\infty}
\left[\dot{D}_k (y-x)-\dot{D}_0 (y-x)\right] V(y) m_n (y,k) dy +
$$
$$
\int_x^{\infty}\left\{\dot{D}_0 (y-x) V(y) \left[m_n (y,k)-m_n (y,0)\right]+
\left[D_k (y-x)-D_0 (y-x)\right] V(y) \dot{m}_n (y,k)\right\}dy +
$$
\beq
\int_x^{\infty}D_0 (y-x) V(y) \left[\dot{m}_n (y,k)-\dot{m}_n (y,0)\right]
 dy.
\label{2.42}
\ene
Moreover, by (\ref{2.23}) and (\ref{2.41})
\beq
\left|\int_0^{\infty}\left[ \dot{D}_k (y-x) -\dot{D}_0 (y-x)\right] V(y)
m_n (y,k) dy \right| \leq C_{x_0} |k|^{\gamma -2}, x \geq x_0.
\label{2.43}
\ene
By (\ref{2.29}) with $\gamma =2$
\beq
\left|m_{1,n}(x,k)-m_{1,n}(x,0)\right|=\left|\int_0^k \dot{m}_{1,n}(x,s)ds
\right| \leq C_{x_0}\, |k|, x \geq x_0,
\label{2.44}
\ene
and then by (\ref{2.15})
\beq
\left|\int_x^{\infty} \dot{D}_0 (y-x) V(y) \left[m_n (y,k)-m_n (y,0)\right]dy
\right| \leq C_{x_0}\, |k|,\, x \geq x_0.
\label{2.45}
\ene
Moreover, by (\ref{2.6})
\beq
|D_k (y)- D_0 (y)| \leq |k||y|^2,
\label{2.46}
\ene
and it follows from (\ref{2.29}) with $\gamma =2$ that
\beq
\left|\int_x^{\infty}\left[D_k (y-x)-D_0 (y-x)\right] V(y) \dot{m}_n (y,k)dk
\right|\leq C_{x_0} |k|, x \geq x_0.
\label{2.47}
\ene
Then we obtain from (\ref{2.21}), (\ref{2.42}), (\ref{2.43}), (\ref{2.45}) and
(\ref{2.47}) that for $|k| \leq 1$:
$$
\left|\dot{m}_{n+1}(x,k)-\dot{m}_{n+1}(x,0)\right|\leq C_{x_0}|k|^{\gamma-2}+
$$
\beq
\int_x^{\infty} (y-x) |V(y)| \left|\dot{m}_n (y,k)-\dot{m}_n (y,0)\right|dy,
x \geq x_0.
\label{2.48}
\ene
But since $m_0 (x,k) \equiv 1$ it follows from (\ref{2.48}) with $n=0$ that
\beq
\left|\dot{m}_{1,1}(x,k) -\dot{m}_{1,1}(x,0)\right|\leq C_{x_0} |k|^{\gamma-2}.
\label{2.49}
\ene
Iterating (\ref{2.48}) $n$ more times we prove
that
\beq
\left|\dot{m}_{1,n+1}(x,k)-\dot{m}_{1,n+1}(x,0)\right|\leq C_{x_0}
 |k|^{\gamma-2}\left(1+\sum_{l=1}^{n} \frac{(q(x))^l}{l!}\right),
\label{2.50}
\ene
with $q(x)$ as in (\ref{2.28}) and taking the limit as
$n\rightarrow \infty$ we have that
\beq
\left|\dot{m}_1 (x,k)- \dot{m}_1 (x,0)\right| \leq C_{x_0} |k|^{\gamma -2}
e^{q(x)}, x \geq x_0 ,
\label{2.51}
\ene
and this proves (\ref{2.37}). Equation (\ref{2.38}) follows similarly.

\begin{corollary}
 Suppose that $V \in L^1_{\gamma}$ for some $2 \leq \gamma \leq 3$. Then for
 every $x_0 \in \ER$ there is a constant $C_{x_0}$ such that
 \beq
 \left| \dot{\acute{m}}_1 (x,k) -\dot{\acute{m}}_1 (x,0)\right| \leq
 C_{x_0}\, |k|^{\gamma -2}, x \geq x_0 ,
 \label{2.52}
 \ene
 \beq
 \left| \dot{\acute{m}}_2 (x,k)-\dot{\acute{m}}_2 (x,0)\right|\leq C_{x_0}
 \,|k|^{\gamma -2}, x \leq x_0 .
 \label{2.53}
 \ene
 \end{corollary}

 \noindent {\it Proof:} \, We give the proof of (\ref{2.52}). Equation
 (\ref{2.53}) follows in a similar way. By (\ref{2.35})
 $$
 \dot{\acute{m}}_1 (x,k) - \dot{\acute{m}}_1 (x,0)= -\int_x^{\infty}
 dy\, \left[ e^{2ik(y-x)}-1\right] V(y) \left\{2i(y-x) m_1 (y,k)+
 \dot{m}_1 (y,k) \right\}-
  $$
  \beq
  \int_x^{\infty}dy\, V(y) \left[ 2i (y-x) \left(m_1 (y,k) -m_1 (y,0)\right)+
  \dot{m}_1 (y,k) -\dot{m}_1 (y,0)\right] dy.
  \label{2.54}
  \ene
  Then by (\ref{2.7}), (\ref{2.12})with $\gamma =2$ and (\ref{2.37})
\beq
\left|\dot{\acute{m}}_1 (x,k)-\dot{\acute{m}}_1 (x,0)\right| \leq C_{x_0}
\, |k|^{\gamma -2}, x \geq x_0 .
\label{2.55}
\ene

\bull

The Jost solutions, $f_j (x,k), j=1,2,$ are independent solutions to
(\ref{2.1}) for $k\neq 0$ and there are unique functions $T(k)$ and
$R_j (k), j=1,2,$ such that \cite{1}

\beq
f_2 (x,k)= \frac{R_1 (k)}{T(k)} f_1 (x,k) +\frac{1}{T(k)} f_1 (x,-k),
\label{2.56}
\ene
\beq
f_1 (x,k)= \frac{R_2 (k)}{T(k)} f_2 (x,k)+ \frac{1}{T(k)} f_2 (x,-k),
\label{2.57}
\ene
for $k \in \ER \setminus 0$. The function $T(k) f_1 (x,k)$ describes the
scattering from left to right of a plane wave $e^{ikx}$ and $T(k) f_2 (x,k)$
describes the scattering from right to left of a plane wave $e^{-ikx}$. The
function $T(k)$ is the transmission coefficient , $R_2 (k)$ is the reflection
coefficient from left to right and $R_1 (k)$ is the reflection coefficient
from right to left. The relations (\ref{2.56}) and (\ref{2.57}) are expressed
as follows in terms of the $m_j (x,k), j=1,2,$
\beq
T(k) m_2 (x,k)=R_1 (k) e^{2ikx} m_1 (x,k) +m_1 (x,-k),
\label{2.58}
\ene
\beq
T(k) m_1 (x,k)= R_2 (k) e^{-2ik} m_2 (x,k) +m_2 (x,-k).
\label{2.59}
\ene
Moreover, $T(k)$ is meromorphic for $\Im k > 0$ with a finite number of
simple poles,\linebreak
$i\beta_N , i\beta_{N-1},\cdots  ,i\beta_1,\, \beta_j > 0,\,j=1,2,
\cdots ,N,$ on the imaginary axis. The numbers, $-\beta^2_N ,\linebreak
 -\beta^2_{N-1},
\cdots ,-\beta^2_1 ,$ are the simple eigenvalues of $H$. Furthermore, $T(k)$
is continuous in $\Im k \geq 0, k\neq i\beta_1 ,i\beta_2, \cdots i\beta_N $
and $T(k)\neq 0$ for $k \neq 0$. the $R_j (k),j=1,2,$ are continuous for $k
\in \ER$. Moreover, the following formulas hold \cite{1}

\beq
\frac{1}{T(k)}= \frac{1}{2ik} [f_1 (x,k), f_2 (x,k)]=1-\frac{1}{2ik}
\int_{-\infty}^{\infty} V(y)\, m_j (y,k)\, dk,\, j=1,2.
\label{2.60}
\ene
 \beq
 \frac{R_1 (k)}{T(k)}=\frac{1}{2ik} [f_2 (x,k), f_1 (x, -k)]=\frac{1}{2ik}
 \int_{-\infty}^{\infty} e^{-2iky}\,V(y)\, m_2 (y,k)\,dy,
 \label{2.61}
 \ene
 \beq
 \frac{R_2 (k)}{T(k)}=\frac{1}{2ik} [f_2 (x,-k) ,f_1 (x,k)]=\frac{1}{2ik}
 \int_{-\infty}^{\infty} e^{2iky} V(y)\, m_1 (y,k)\, dy.
 \label{2.62}
 \ene
 Furthermore,
 \beq
 T(k)=1+O\left(\frac{1}{|k|}\right),\, |k|\rightarrow \infty , \Im k \geq 0,
 \label{2.63}
 \ene
 \beq
 R_j (k)= O\left(\frac{1}{|k|}\right), |k| \rightarrow \infty , k \in \ER,
 \label{2.64}
 \ene
 and
 \beq
 |T(k)|^2 +|R_j (k)|^2 =1, \, j=1,2, \, k \in \ER .
 \label{2.65}
 \ene
 The behaviour as $ k \rightarrow 0$ is as follows:

 \noindent (a) In the {\it generic case}
 \beq
 T(k)= \alpha k +o(k),\, \alpha \neq 0 ,k \rightarrow 0 ,\Im k \geq 0,
 \label{2.66}
 \ene
 and $R_1 (0)=R_2 (0)= -1$.

 \noindent (b) In the {\it exceptional case}
 \beq
 T(k)= \frac{2a}{1+a^2}+o(1),\, k \rightarrow 0 , \Im k \geq 0,
 \label{2.67}
 \ene
 \beq
 R_1 (k)= \frac{1-a^2}{1+a^2} +o(1), \, k \rightarrow 0 ,\, k\in \ER ,
 \label{2.68}
 \ene
 \beq
 R_2 (k)=\frac{a^2 -1}{1+a^2}+o(1),\, k \rightarrow 0,\, k \in \ER ,
 \label{2.69}
 \ene
 where $a= \lim_{x \rightarrow -\infty}\, f_1 (x,0)\neq 0.$
 For the results above about $T(k)$ and $R_j (k), \,j=1,2,$ see \cite{1},
 \cite{2} and \cite{3}. In particular for the continuity of $T(k)$ and of
 $R_j (k)$  as $k \rightarrow 0$ in the exceptional case for $V \in L^1_1$
 see \cite{3}.

\begin{theorem}Assume that $V\in L^1_{\gamma}$.

\noindent (a)If $V$ is {\rm generic} and $1 \leq \gamma \leq 2$, then
\beq
|\dot{T}(k)|\leq C (1+|k|)^{-1} ,\, \Im k \geq 0,
\label{2.70}
\ene

\beq
R_j (k_1 )-R_j (k_2 )=\cases{ o\left(|k_1 -k_2 |^{\gamma -1}\right),&
$1\leq \gamma <2$, \cr\cr
O(|k_1 -k_2 |), & $\gamma =2$, }
\label{2.72}
\ene
as $ k_1 -k_2 \rightarrow 0$.

\noindent (b) If $V$ is {\rm exceptional} and $ 2 \leq \gamma \leq 3$, then
\beq
|\dot{T}(k)|\leq C \frac{|k|^{\gamma -3}}{(1+|k|)^{\gamma -2}},
\label{2.73}
\ene
\beq
T(k)-T(0)=O(|k|), k \rightarrow 0,
\label{2.74}
\ene

\beq
R_j (k)-R_j (0)=O(|k|),\, k \rightarrow 0,\, j=1,2.
\label{2.76}
\ene
Moreover, if $\gamma > 2$
\beq
R_j (k_1 )-R_j (k_2 )= O\left(|k_1 -k_2 |^{\gamma -2}\right), \,
k_1 -k_2 \rightarrow 0.
\label{2.77}
\ene
\end{theorem}

\noindent{\it Proof:}\, It follows from (\ref{2.7}) and (\ref{2.34}) that
\beq
\left|\acute{m}_1 (x,k)\right| \leq C,\, x\in \ER ,\, \Im k \geq 0.
\label{2.78}
\ene
We similarly prove that
\beq
\left|\acute{m}_2 (x,k)\right| \leq C, \, x\in \ER , \, \Im k \geq 0.
\label{2.79}
\ene
Then (\ref{2.70}) follows from (\ref{2.7}), (\ref{2.8}), (\ref{2.12}),
(\ref{2.13}), (\ref{2.32}), (\ref{2.33}), the first equality in
(\ref{2.60}), (\ref{2.63}),
(\ref{2.66}), (\ref{2.78}) and (\ref{2.79}).

If follows from (\ref{2.19}) that
$$
m_{1,n+1}(x,k_1 )-m_{1,n+1}(x, k_2 )=f_n (x,k_1 ,k_2 )+
\int_x^{\infty}D_{k_2}(y-x) V(y)
$$
\beq
\left[m_{1,n}(y,k_1 )-m_{1,n}(y,k_2 ) \right] dy,
\label{2.80}
\ene
where
\beq
f_n (x,k_1 ,k_2 ):= \int_x^{\infty} \left[D_{k_1}(y-x)-D_{k_2}(y-x)\right]
V(y) m_{1,n}(y,k_1 ) \, dy.
\label{2.81}
\ene
Moreover, by (\ref{2.6})
\beq
\left|D_{k_1}(x)-D_{k_2}(x)\right|\leq 2
\frac{|k_1 -k_2 ||x|}{1+ |k_1 -k_2 ||x|} |x|.
\label{2.82}
\ene
Then by (\ref{2.23}) for $x \geq 0$
\beq
\left|f_n (x,k_1 ,k_2 )\right| \leq f_{\gamma} (k_1 -k_2 ),
\label{2.83}
\ene
where for $1 \leq \gamma \leq 2$
\beq
f_{\gamma}(k)= C\, |k|^{\gamma-1} \int_0^{\infty} \, y^{\gamma} |V(y)|
\left(\frac{|k|y}{1+|k|y}\right)^{2-\gamma}\, dy.
\label{2.84}
\ene
Note that as $k \rightarrow 0$
\beq
f_{\gamma}(k)=\cases{ o\left(|k|^{\gamma -1}\right),& $1 \leq \gamma < 2$,\cr\cr
O\left( |k|\right),& $\gamma =2$.}
\label{2.85}
\ene
Since the function: $\lambda \rightarrow  |k| \lambda (1+|k| \lambda )^{-1}$
is an increasing function
of $\lambda $, for $\lambda \geq 0 $, we have that (see (\ref{2.7}) and
(\ref{2.82})) for all $x\in \ER $
$$
\int_0^{\infty}\left|D_{k_1}(y-x)-D_{k_2}(y-x)\right| |V(y)\, m_{1,n}(y,k_2 )|
\,dy
\leq C \int_0^{\infty} \frac{|k_1 -k_2 |(|x|+|y|)}{1+|k_1 -k_2 | (|x|+|y|)}
$$
$$
(|x|+|y|) |V(y)| dy\leq C\, \frac{|k_1 -k_2 ||x|^2}{1+|k_1 -k_2 ||x|}
+C\, \int_0^{\infty}\frac{|k_1 -k_2 ||y|^2}{1+|k_1 -k_2 ||y|} dy
$$
\beq
\leq C\left[\frac{|k_1 -k_2 ||x|}{1+|k_1 -k_2 ||x|}+f_{\gamma}(k_1-k_2 ).
\right] (1+|x|).
\label{2.86}
\ene
Furthermore, for $x \leq 0$ (see (\ref{2.7}) and (\ref{2.82}))
$$
\int_x^0\left|D_{k_1}(y-x)-D_{k_2}(y-x)\right||V(y)\,m_{1,n}(y,k_2 )|
dy \leq
$$
\beq \int_x^0 \frac{|k_1 -k_2 |\, (|x|+|y|)^2}{1+|k_1 -k_2 |(|x|+|y|)}\,
|V(y)|\,(1+|y|) dy \leq C \frac{|k_1 -k_2 ||x|^2}{1+|k_1 -k_2 ||x|}.
\label{2.87}
\ene

By (\ref{2.86}) and (\ref{2.87}) for $x \leq 0$
\beq
\left|f_n (x,k_1 ,k_2 )\right|\leq g_{\gamma}(x,k_1 -k_2 )
\label{2.88}
\ene
where
\beq
g_{\gamma}(x,k):=C\, \left[ \frac{|k||x|}{1+|k||x|}+f_{\gamma}(k)\right]
(1+|x|).
\label{2.89}
\ene
By (\ref{2.80}) and (\ref{2.83}) we have that for $x \geq 0$
\beq
\left|m_{1,n+1}(x,k_1 )-m_{1,n+1}(x,k_2 )\right|\leq f_{\gamma}(x, k_1 -k_2 )
+\int_x^{\infty} \left|m_{1,n}(y,k_1 )-m_{1,n}(y,k_2 )\right|\,y\, |V(y)| dy.
\label{2.90}
\ene
Since $m_{1,0}(x,k) \equiv 1$, it follows from (\ref{2.80}) and (\ref{2.83})
that
\beq
\left|m_{1,1}(x, k_1 )- m_{1,1}(x, k_2 )\right|\leq f_{\gamma}(x, k_1 -k_2 ),\,
x \geq 0.
\label{2.91}
\ene
Then iterating (\ref{2.90}) we prove that
\beq
\left|m_1 (x,k_1 )-m_1 (x,k_2 )\right| \leq f_{\gamma}(x, k_1 -k_2 )\,
e^{\left(\int_x^{\infty} y \, |V(y)|\, dy\right)},\, x \geq 0.
\label{2.92}
\ene
Moreover, taking the limit as $n \rightarrow \infty$ in
(\ref{2.80}) and using (\ref{2.21}), (\ref{2.88}) and (\ref{2.92})
we obtain that for $x \leq 0$
\beq
\left|m_1 (x,k_1 )-m_1 (x,k_2 )\right|\leq g_{\gamma}(x,k_1 -k_2 )+
 \int_x^0 (|x|+|y|) |V(y)| \left|m_1 (y,k_1 )-m_1 (y,k_2 )\right| dy,
\label{2.93}
\ene
where in the right--hand side of (\ref{2.89}) we take a constant $C$ large
enough. Let us denote
\beq
h(x,k_1 ,k_2 ):= \frac{\left|m_1 (x,k_1 )- m_1 (x,k_2 )\right|}
{g_{\gamma}(x, k_1 -k_2 )}.
\label{2.94}
\ene
Then it follows from (\ref{2.93}) that for $x \leq 0$
\beq
h(x,k_1 ,k_2 )\leq 1 + \int_0^x (1+|y|) |V(y)| h(y, k_1 ,k_2 ) dy,
\label{2.95}
\ene
where we used that $ g_{\gamma}(x,k) /(1+|x|)$ is an increasing function
of $|x|$.
By (\ref{2.95}) and Gronwall's inequality (see page 204 of \cite{7}) we have
that
\beq
h(x, k_1 ,k_2 )\leq e^{\int_0^{\infty} (1+|y|)|V(y)| dy}
\label{2.96}
\ene
 and then taking in (\ref{2.89}) $C$ large enough we obtain that
 \beq
 \left|m_1 (x,k_1 )-m_1 (x,k_2 )\right|\leq g_{\gamma}(x,k_1 -k_2 ).
 \label{2.97}
 \ene
 We similarly prove that
 \beq
 \left|m_2 (x,k_1 )-m_2 (x,k_2 )\right|\leq g_{\gamma}(x, k_1 -k_2 ).
 \label{2.98}
 \ene
 Note that in the proof of (\ref{2.97}), (\ref{2.98}) we only used that
 $V \in L^1_{\gamma},\, 1 \leq \gamma \leq 2$. We now prove (\ref{2.72}).
 It follows from (\ref{2.58}) that
 $$
 R_1 (k_1 )-R_1 (k_2 )=\left( m_1 (x,k_2 )\right)^{-1} \left[ e^{-2ik_1 x}
 T(k_1 )\, m_2 (x,k_1 ) -e^{-2ik_2 x} T(k_2 )\, m_2 (x,k_2 ) +\right.
 $$
 \beq \left.
 e^{-2ik_2 x}
 m_1 (x,-k_2)
 -e^{-2ik_1 x} m_1 (x, -k_1 ) +R_1 (k_1 ) (m_1 (x,k_2 )-m_1 (x,k_1 )) \right].
 \label{2.99}
 \ene
 Then by (\ref{2.4}) and (\ref{2.7}) there is an $x_0 \in \ER$
 such that
 \beq
 |m_1 (x,k)|\geq \frac{1}{2},\, x \geq x_0 , \, k \in \ER.
 \label{2.100}
 \ene
 Then (\ref{2.72}) with $j=1$ follows from (\ref{2.70}), (\ref{2.97})  and
 (\ref{2.98}) taking in (\ref{2.99}) any $x \geq x_0$. Equation (\ref{2.72})
 with $j=2$ is proven in a similar way.
Equation (\ref{2.73}) follows from (\ref{2.7}), (\ref{2.8}), (\ref{2.12}),
(\ref{2.13}), (\ref{2.32}), (\ref{2.33}), (\ref{2.37}),
(\ref{2.38}), (\ref{2.52}), (\ref{2.53}), the first equality in the
right--hand side of (\ref{2.60}) and (\ref{2.65}) and noting that if
$V \in L^1_2 $
\beq
 [f_1 (x,k), f_2 (x,k)]=ik\frac{1+a^2}{a}+O\left(k^2 \right), k \rightarrow 0.
 \label{2.106}
 \ene
 Equation (\ref{2.106}) is proven by the argument given in \cite{3} to prove
 that
 \beq
 [f_1 (x,k),f_2 (x,k)]=ik \frac{1+a^2}{a}+o(k), k \rightarrow 0,
 \label{2.107}
 \ene
 in the case when $V \in L^1_1 $. The fact that in (\ref{2.106}) we have
 $O\left(k^2 \right)$ instead of $o(k)$ follows because we assume that
 $V \in  L^1_{\gamma} , \gamma \geq 2 $ (see (\ref{2.12}) and (\ref{2.13})).
  Equation (\ref{2.74}) follows from the first equality
  in the right--hand side of
 (\ref{2.60}) and by (\ref{2.106}). Also (\ref{2.76}) follows from the
 first equality in the right--hand side of (\ref{2.61}) and (\ref{2.62})and
 observing that
 \beq
 \left[ f_1 (x,k), f_2 (x, -k)\right] =-ik \frac{a^2 -1}{a}
 +O\left(k^2 \right), \, k \rightarrow 0.
 \label{2.101}
 \ene
Equation (\ref{2.101}) is proven as (\ref{2.106}). It follows from (\ref{2.73})
that
 \beq
 T(k_1) -T(k_2 )=O\left(|k_1 -k_2 |^{\gamma -2}\right), k_1 -k_2 \rightarrow 0.
 \label{2.108}
 \ene
 Then (\ref{2.77}) with $j=1$ follows from (\ref{2.97}), (\ref{2.98}),
 (\ref{2.99}) and (\ref{2.108}). Equation (\ref{2.77}) with $j=2$ is proven
 in the same way.

 \bull

The results on the spectral theorem for $H$ that we state below follow from
the Weyl--Kodaira--Titchmarsch theory. See for example \cite{1}. For a
version of the Weyl--Kodaira-- Titchmarsch theory adapted to our situation
see Appendix 1 of \cite{4} and also the proof of Theorem 6.1 in page 78 of
\cite{4}. Let us denote for any $k \in \ER$
\beq
\Psi_+ (x,k):= \cases{ \frac{1}{\sqrt{2 \pi }} T(k) f_1 (x,k),& $ k \geq 0$,\cr\cr
\frac{1}{\sqrt{2\pi}} T(-k) f_2 (x,-k),& $k < 0$,}
\label{2.109}
\ene
and $\Psi_- (x,-k) := \overline{\Psi_+ (x,k)}$. Let ${\cal H}_{ac}(H)$ be the
subspace of absolute continuity of $H$. Then the following limits
\beq
\hat{\phi}_{\pm}(k):= s-\lim_{N \rightarrow \infty}\int_{-N}^{N}
\overline{\Psi_{\pm}(x,k)} \,\phi (x) \, dx
\label{2.110}
\ene
exist in the strong topology in $L^2$ for every $\phi \in L^2 $ and the
operators
\beq
\left(F_{\pm} \phi \right)(k):= \hat{\phi}_{\pm}(k)
\label{2.111}
\ene
are unitary operators from ${\cal H}_{ac}(H)$ onto $L^2$. Moreover, the
${F^{\ast}}_{\pm}$ are given by
\beq
\left({F^{\ast}}_{\pm} \phi \right)(x)= s-\lim_{N\rightarrow \infty}
\int_{-N}^{N} \Psi_{\pm}(x,k) \,\phi (k) \, dk,
\label{2.112}
\ene
where the limits exist in the strong topology in $L^2$. Furthermore, the
operators ${F^{\ast}}_{\pm }F_{\pm}$ are the orthogonal projection onto
${\cal H}_{ac}(H)$. For each eigenvalue of $H$, let $\Psi_j , j=1,2,\cdots ,N$
be the corresponding eigenfunction normalized to one, i.e.
$\|\Psi_j \|_{L^2}=1$. The operators:
\beq
F_j \phi := (\phi ,\Psi_j ) \Psi_j, j=1,2,\cdots , N,
\label{2.113}
\ene
are unitary from the eigenspace generated by $\Psi_j$ onto $C$. The following
operators
\beq
F^{\pm}= F_{\pm} \oplus^N_{j=1} F_j ,
\label{2.114}
\ene
are unitary from $L^2$ onto $L^2 \oplus^N_{j=1} C$ and for any $\phi \in D(H)$
\beq
F^{\pm} H \phi= \left\{ k^2 (F_{\pm} \phi )(k) , -\beta^2_1 F_1 \phi ,
\cdots , - \beta^2_N F_N \phi \right\}.
\label{2.115}
\ene
Moreover, for any bounded Borel function , $ \Phi$, defined on $\ER$
\beq
F^{\pm} \Phi (H) \phi = \left\{ \Phi (k^2) (F_{\pm}\phi )(k), \Phi(-\beta^2_1
)F_1 \phi ,\cdots , \Phi (-\beta^2_N ) F_N \phi \right\}.
\label{2.116}
\ene
The projector, $P_p$, onto the subspace of $L^2$ generated by the eigenvectors
of $H$ is given by
\beq
P_p \phi:= \sum_{j=1}^N  (\phi ,\Psi_j ) \Psi_j .
\label{2.117}
\ene
 Since $H$ has no
singular--continuous spectrum the projector onto the continuous subspace of
$H$ is given by: $P_c := I-P_p$. It follows from (\ref{2.116}) that
 \beq
 e^{-it H}P_c = F^{\pm \ast} e^{-ik^2 t} F^{\pm}.
 \label{2.118}
 \ene
 Equation (\ref{2.118}) is the starting point of our proof of the
  $L^1 - L^{\infty} $ estimate (Theorem 1.1). We divide the proof of the
  $L^1 -L^{\infty}$  estimate into a high--energy estimate and a low--energy
  estimate. For this purpose, let $\Phi$ be any continuous and bounded function
  on $\ER$ that has a bounded derivative and such that $\Phi (k) =0$ for
  $|k| \leq k_1 $ and $\Phi (k)=1$ for $|k| \geq k_2 $ for some $0 < k_1  < k_2$.

\begin{lemma}(The high--energy estimate). Suppose that $V \in L^1_1$. Then
$e^{-itH} \Phi (H)P_c $ extends to a bounded operator from $L^1$ to $L^{\infty}$ and
there is a constant $C$ such that
\beq
\left\| e^{-itH} \Phi (H) P_c \right\|_{{\cal B}\left( L^1 ,L^{\infty}\right)} \leq
 \frac{C}{\sqrt{t}}, \, t > 0.
 \label{2.119}
 \ene
 \end{lemma}
 \noindent {\it Proof:}\,Let us take $\chi \in C^{\infty}, \, \chi (k)=1,
 |k| \leq 1$ and $\chi (k)=0, \, k \geq 2$, and let us denote $ \chi_n (k)=
 \chi (k/n), n=1,2,\cdots $. Then it follows from (\ref{2.118}) that for any
 $f, g \in L^1 \cap L^2 $:
\beq
 \left(e^{-itH}\Phi (H) P_c f, g \right)= \lim_{n \rightarrow \infty}
 \left( e^{-itH} \Phi (H) \chi_n (H) P_c f, g \right)=
\lim_{n \rightarrow \infty}\int dx \, dy \Phi_{t,n}(x,y) f(x) \overline{g(y)},
\label{2.120}
\ene
where
\beq
\Phi_{t,n}(x,y):=\int_{-\infty}^{\infty} e^{-ik^2 t} \chi_n (k^2 ) \Phi (k^2 )
\overline{\Psi_+ (x,k)} \Psi_+ (y,k) dk.
\label{2.121}
\ene
We have that,
\beq
\Phi_{t,n}(x,y) = \Phi^{(0)}_{t,n} (x,y)+\Phi^{(1)}_{t,n}(x,y)
+\Phi^{(+)}_{t,n}(x,y)+\Phi^{(-)}_{t,n}(x,y),
\label{2.122}
\ene
where
\beq
\Phi^{(0)}_{t,n}(x,y):= \int_{-\infty}^{\infty} e^{-ik^2 t} \,
\frac{e^{-ik(x-y)}}{2 \pi} \chi_n (k^2 ) dk,
\label{2.123}
\ene
\beq
\Phi^{(1)}_{t,n}(x,y) :=\int_{-\infty}^{\infty} e^{-ik^2 t} \,
\frac{e^{-ik(x-y)}}{2 \pi}  \chi_n (k^2 ) (\Phi (k^2 )-1) dk,
\label{2.124}
\ene
\beq
\Phi^{(+)}_{t,n}(x,y):= \int_0^{\infty}e^{-ik^2 t}\,
\frac{e^{-i(x-y)}}{2 \pi } \chi_n (k^2 ) m_+ (x,y,k) dk,
\label{2.125}
\ene
\beq
\Phi^{(-)}_{t,n}(x,y):= \int_{-\infty}^0 e^{-ik^2 t}\,
\frac{e^{-ik(x-y)}}{2\pi } \chi_n (k^2 ) m_- (x,y,k) dk,
\label{2.126}
\ene
with
$$
m_{\pm}(x,y,k):= \Phi (k^2 )\left[ \overline{(T(k) m_{j(\pm )}(x,k)-1)} T(k)
m_{j(\pm)}(y,k) + \right.
$$
\beq
\left. T(k) m_{j(\pm)} (y,k) -1 \right],
\, \pm k \geq 0,
\label{2.127}
\ene
where $j(+)=1$ and $j(-)=2$.
Since the inverse Fourier transform of $\frac{1}{\sqrt{2 \pi} }e^{-ik^2 t}$
 is
 \beq
\Phi^{(0)}_t (x):= \frac{1}{\sqrt{4 \pi i t}}\, e^{ix^2 / 4t}
\label{2.127b}
\ene
it follows that
\beq
\lim_{n \rightarrow \infty} \int dx \, dy \, \Phi^{(0)}_{t,n}(x,y) f(x)
\overline{g(y)}= \int dx \, dy\, \Phi^{(0)}_t (x,y) f(x)
\overline{g(y)}.
\label{2.128}
\ene
Changing the coordinates of integration in (\ref{2.124}) to $p=k-k_0$ where
$k_0 = (y-x)/2t$ we obtain that
$$
\Phi^{(1)}_{t,n} (x,y)=\frac{1}{2 \pi}  e^{i(x-y)^2 /4t}
\int_{-\infty}^{\infty} dp e^{-ip^2 t}
\chi_n \left((p+k_0 )^2 \right) \left(\Phi\left((p+k_0 )^2 \right)-1 \right)=
$$
\beq
\frac{1}{2 \pi \sqrt{2it}}\,  e^{i(x-y)^2 /4t} \int_{-\infty}^{\infty}
dp \, e^{i \rho^2 /4t} \, \hat{h}_n (\rho ),
\label{2.131}
\ene
where in the second equality we used the Plancherel theorem and
$\hat{h}_n (\rho )$ is the Fourier transform of the function
$ h_n (\rho )$  defined as
follows
\beq
h_n (\rho ):= \overline{ \chi_n \left((p+k_0 )^2 \right)
\left(\Phi \left((p+k_0 )^2\right)-1 \right)}.
\label{2.132}
\ene
Since,
\beq
\left\| \hat{h}_n \right\|_{L^1} \leq C \left\|h_n \right\|_{W_1} \leq C
\left\|\Phi (p^2 )-1 \right\|_{W_1},
\label{2.133}
\ene
we have that
\beq
\left| \Phi^{(1)}_{t,n}(x,y) \right| \leq \frac{C}{\sqrt{t}}.
\label{2.134}
\ene
Let us denote $h(p):= \overline{ \Phi \left((p+k_0 )^2 \right)-1}$. Then
since $\hat{h}_n (p)$ converges to $\hat{h}(p)$ in the $L^1$ norm, it
follows  from (\ref{2.131})
and the dominated convergence theorem  that
\beq
\lim_{n \rightarrow \infty} \Phi^{(1)}_{t,n} (x,y) =\Phi^{(1)}_t (x,y):=
\frac{1}{2 \pi \sqrt{2it}} \,e^{i(x-y)^2 /4t} \int_{-\infty}^{\infty}
e^{i \rho^2 /4t} \, \hat{h}(\rho )d\rho,
\label{2.135}
\ene
and that
\beq
\left| \Phi^{(1)}_t (x,y)\right| \leq \frac{C}{\sqrt{t}}, \, x,y \in \ER.
\label{2.136}
\ene
Using the dominated convergence theorem again we prove that
\beq
\lim_{n \rightarrow \infty}\int dx \, dy\, \Phi^{(1)}_{t,n}(x,y) f(x)
\overline{g(y)}= \int dx \, dy \,\Phi^{(1)}_t (x,y) f(x) \overline{g(y)}.
\label{2.137}
\ene
We denote
\beq
m_{+,e} (x,y,k):=\cases{ m_+ (x,y,k),& $k \geq 0$, \cr\cr
0,& $k < 0 $.}
\label{2.138}
\ene
Then since $\Phi (k^2 )=0$ for $|k| \leq \sqrt{k_1}$ and $\Phi (k^2 )=1$
for $|k| \geq \sqrt{k_2}$, it follows from (\ref{2.7}), (\ref{2.12}),
 (\ref{2.63}), (\ref{2.70}) and (\ref{2.127}) that for some constant $C$
 \beq
 \left\|m_{+,e}(x,y,\cdot )\right\|_{W_1} \leq C, \,\,x, y \geq 0.
 \label{2.139}
 \ene
 Then, as in the case of $\Phi^{(1)}_{t,n}$ we prove that
 \beq
 \left|\Phi^{(+)}_{t,n}(x,y)\right| \leq \frac{C}{\sqrt{t}}, \,\,x,y \geq 0,
  t > 0.
  \label{2.140}
  \ene
  and that
  \beq
  \lim_{n \rightarrow \infty} \Phi^{(+)}_{t,n}(x,y)=\Phi^{(+)}_t (x,y):=
  \frac{1}{2 \pi \sqrt{2it}}\int_{-\infty}^{\infty}
  e^{i\rho^2 /4t} \,\tilde{m}_{+,e}(x,y,\rho )d \rho,
  \label{2.141}
  \ene
  where $\tilde{m}_{+,e}(x,y,\rho )$ is the Fourier transform of
  $m_{+,e}(x,y,k+k_0 )$, and that
  \beq
  \left|\Phi^{(+)}_{t}(x,y)\right| \leq \frac{C}{\sqrt{t}},\,\, x,y \geq 0,
  t > 0.
  \label{2.142}
  \ene
  Using (\ref{2.58}) we write (\ref{2.126}) as follows
  \beq
  \Phi^{(-)}_{t,n} (x,y)=\sum_{j=2}^5 \Phi^{(j)}_{t,n}(x,y),
  \label{2.143}
  \ene
  where
  \beq
  \Phi^{(j)}_{t,n}(x,y):=\int_{-\infty}^0 e^{-ik^2 t}
  \frac{e^{-i k (lx-ry)}}{2\pi}\, \chi_n (k^2)\,m_j (x,y,k) dk,
  \label{2.144}
  \ene
 where for $j=2,\, l=r=3$, for $j=3, \, l=3, r=1$, for $j=4, \, l=1, r=3$,
 and for $j=5,\, l=r=1$. Moreover, (recall that
 $m_j (x,-k)=\overline{m_j (x,k)}$)
 \beq
 m_2 (x,y,k):= \Phi (k^2 )\left[|R_1 (k)|^2 \overline{m_1 (x,k)} m_1 (y,k)
 \right],
 \label{2.145}
 \ene
 \beq
 m_3 (x,y,k):= \Phi (k^2 ) \overline{R_1 (k^2 ) m_1 (x,k) m_1 (y,k)},
 \label{2.146}
 \ene
 \beq
 m_4 (x,y,k):= \Phi(k^2 )R_1 (k) (\overline{m_1 (x,k)}-1) m_1 (y,k),
 \label{2.147}
 \ene
and
\beq
m_5 (x,y,k):= \Phi (k^2 )\overline{(m_1 (x,k)-1) m_1 (y,k)}.
\label{2.148}
\ene
Then as in the case of $\Phi^{(+)}_{t,n}$ we prove that
\beq
\left|\Phi^{(-)}_{t,n}(x,y)\right|\leq \frac{C}{\sqrt{t}},\, x,y \geq 0,
t > 0,
\label{2.149}
\ene
and that
\beq
\lim_{n\rightarrow \infty}\Phi^{(-)}_{t,n}(x,y)=\Phi^{(-)}_t (x,y),
\, x,y \geq 0, t > 0,
\label{2.150}
\ene
where
\beq
\Phi^{(-)}_t (x,y)=\sum _{j=2}^5 \Phi^{(j)}_t (x,y),
\label{2.151}
\ene
with
\beq
\Phi^{(j)}_t (x,y):= \frac{1}{2 \pi \sqrt{2t}}\int_{-\infty}^{\infty}
e^{i\rho^2 /4t} \tilde{m}_j (x,y,\rho )d\rho,
\label{2.152}
\ene
with $\tilde{m}_j (x,y,\rho )$ the Fourier transform of
$m_j (x,y, p+(ry-lx)/2t)$. We also have that

\beq
\left|\Phi^{(-)}_t (x,y)\right|\leq \frac{C}{\sqrt{t}},\, x,y \geq 0, t > 0.
\label{2.153}
\ene
By the same argument as above and using also (\ref{2.59}) we prove that for
$( x \geq 0 , y \leq 0), (x \leq 0 ,y \geq 0 )$ and $ (x \leq 0 ,y \leq 0)$
\beq
\left|\Phi^{(\pm)}_{t,n} (x,y)\right| \leq \frac{C}{\sqrt{t}},\, t > 0,
\label{2.154}
\ene
and that
\beq
\lim_{n \rightarrow \infty}\Phi^{(\pm)}_{t,n}(x,y)= \Phi^{(\pm)}_t (x,y),
\label{2.155}
\ene
for  functions $\Phi^{(\pm)}_t (x,y)$ that satisfy
\beq
\left|\Phi^{(\pm)}_t (x,y)\right| \leq \frac{C}{\sqrt{t}},\, t > 0.
\label{2.156}
\ene
We can explicitly compute $\Phi^{\pm}_t (x,y)$ as in the case
$(x \geq 0 ,y \geq 0)$.Then (\ref{2.154}), (\ref{2.155}) and (\ref{2.156})
hold for all $x,y \in \ER$ and using (\ref{2.120}), (\ref{2.122}),
(\ref{2.128}), (\ref{2.134}), (\ref{2.137}), \linebreak (\ref{2.154} )
and (\ref{2.155})
we prove that
\beq
\left(e^{-itH}\Phi(H) P_c f, g \right)=\int dx \, dy \left[ \Phi^{(0)}_t (x,y)
+\Phi^{(1)}_t (x,y)+ \Phi^{(+)}_t (x,y)+ \Phi^{(-)}_t (x,y)\right] f(x)
\overline{g(y)}.
\label{2.157}
\ene
Then by (\ref{2.127b}), (\ref{2.136}) and (\ref{2.156})
\beq
\left|\left(e^{-itH} \Phi (H) P_c f, g \right)\right|\leq \frac{C}{\sqrt{t}}
\|f\|_{L^1} \|g||_{L^1},\, t > 0,
\label{2.158}
\ene
for all $f, g \in L^1 \cap L^2$. By continuity this estimate holds for all
$f, g \in L^1$ and (\ref{2.119}) follows.

\bull

Let $\Psi$ be any function on $C^{\infty}_0 \left(\ER \right)$ such that
$\Psi (k)=1, |k| \leq \delta$, for some $\delta > 0$.

\begin{lemma} (The low--energy estimate).
Suppose that $V \in L^1_{\gamma}$ where in the {\rm generic case} $\gamma > 3/2$
and in the {\rm exceptional case} $ \gamma > 5/2$. Then $e^{-itH} \Psi(H) P_c$
extends to a bounded operator from $L^1$ to $L^{\infty}$ and there is a
constant $C$ such that
\beq
\left\|e^{-itH} \Psi (H) P_c \right\| \leq  \frac{C}{\sqrt{t}},\, t > 0.
\label{2.159}
\ene
\end{lemma}
\noindent {\it Proof :}\, As in the proof of Lemma 2.6 it follows from
(\ref{2.118}) that for all $f, g \in L^1 \cap L^2$
\beq
 \left(e^{-itH} \Psi (H) P_c f, g \right)=\int dx \, dy \, \Phi_t (x,y)
 f(x) \overline{g(y)},
 \label{2.160}
 \ene
 where
 \beq
 \Phi_t (x,y)= \Phi^{(+)}_t (x,y) +\Phi^{(-)}_t (x,y),
 \label{2.161}
 \ene
 with
 \beq
 \Phi^{(+)}_t (x,y):= \int_0^{\infty} e^{-ik^2 t}\,
 \frac{e^{-ik(x-y)}}{2 \pi}\, m_+ (x,y,k) \,dk,
 \label{2.162}
 \ene
 \beq
 \Phi^{(-)}_t (x,y);= \int_{-\infty}^0 e^{-ik^2 t}\,
 \frac{e^{-ik(x-y)}}{2 \pi} \, m_- (x,y,k) \, dk,
 \label{2.163}
 \ene
 and
 \beq
 m_{\pm}(x,y,k):= \Psi (k^2 ) q_{\pm}(x,y,k)
 \label{2.164}
 \ene
 with
 \beq
 q_{\pm}(x,y,k):=\overline{T(k) m_{j(\pm)}(x,k)} T(k) m_{j(\pm)}(y,k), \,
 \pm k > 0,
 \label{2.165}
 \ene
 where $ j(+)=1$ and $j(-)=2$.

 Let us consider first the {\it generic case}. In this case it follows from
 (\ref{2.66}) that \linebreak $m_{\pm}(x,y, 0{\pm})=0$. We denote
 \beq
 m_{+,e}(x,y,k):=\cases{m_+ (x,y,k),& $ k \geq 0$,\cr\cr
 0,& $k < 0$.}
 \label{2.166}
 \ene
 Let us denote by $\omega_{+,x,y}(\rho )$ the modulus of continuity
 of $m_{+,e}(x,y)$, i.e.,
 \beq
 \omega_{+,x,y}(\rho ):= \left\|m_{+,e}(x,y,k+\rho )
 -m_{+,e}(x,y,k)\right\|_{L^2}.
 \label{2.167}
 \ene
 Remark that
 \beq
 \omega_{+,x,y}(\rho )\leq 2 \left\|m_{+,e}(x,y,\cdot )\right\|_{L^2}
 \leq C_{x_0}, \, x, y \geq x_0.
 \label{2.168}
 \ene
 Without lossing generality we can assume that $ \gamma \leq 2$. Then
 by (\ref{2.7}), (\ref{2.12}), (\ref{2.70}), (\ref{2.163}) and
 (\ref{2.164}) for $|\rho |\leq 1$
 \beq
 \omega_{+,x,y}(\rho )\leq C_{x_0}\, |\rho |^{\gamma -1}, x, y \geq x_0.
 \label{2.169}
 \ene
It follows from (\ref{2.168}) and (\ref{2.169}) that for any $0 \leq \alpha <
\gamma -1 $
\beq
\int d \rho |\omega_{+,x,y}(\rho )|^2 \frac{1}{|\rho |^{1+2 \alpha}}  < \infty
\label{2.170}
\ene
and then by Proposition 4 in page 139 of \cite{5}
\beq
\left\|m_{+,e}(x,y,\cdot )\right\|_{W_{\alpha}} \leq C_{\alpha ,x_0},
x, y \geq x_0,
\label{2.171}
\ene
for any $0 < \alpha < \gamma -1$. Let us denote $k_0 = (y-x)/2t$. Then
we prove as in Lemma 2.6 that
(\ref{2.134})
\beq
\Phi^{(+)}_t (x,y)=\frac{1}{2\pi \sqrt{2it}}\, \int_{-\infty}^{\infty}
e^{i\rho^2 /4t}\, \tilde{m}_{+,e}(x,y,\rho )\, d\rho,
\label{2.172}
\ene
with $\tilde{m}_{+,e}(x,y,\rho )$ the Fourier transform of
$m_{+,e}(x,y,k+k_0 )$. But since for $ \frac{1}{2} < \alpha < \gamma -1$
\beq
\left\|\tilde{m}_{+,e}(x,y,\cdot ) \right\|_{L^1} \leq C
\left\|(1+\rho^2 )^{\frac{\alpha} {2}} \tilde{m}_{+,e}(x,y, \cdot )
\right\|_{L^2}= C \left\|m_{+,e}(x,y,\cdot )\right\|_{W_{\alpha}} \leq
C , \, x, y \geq 0 ,
\label{2.173}
\ene
we have that
\beq
\left|\Phi^{(+)}_t (x,y)\right| \leq \frac{C}{\sqrt{t}},\, x, y \geq 0
,  t > 0.
\label{2.174}
\ene
Using (\ref{2.7}), (\ref{2.8}), (\ref{2.12}), (\ref{2.13}),
(\ref{2.58}), (\ref{2.59}), (\ref{2.61}) and (\ref{2.72}) we prove in the
same way that
(\ref{2.174}) holds for $(x \geq 0 , y < 0), (x \leq 0, y \geq 0)$
and $(x \leq 0, y\leq 0)$ and that the same is true for $\Phi^{(-)}_t (x,y)$
(see the proof of Lemma 2.6 for a similar argument). Then we have that
\beq
\left|\Phi_t (x, y)\right| \leq \frac{C}{\sqrt{t}}, \,x, y \in \ER,\, t > 0.
\label{2.175}
\ene
Equation (\ref{2.159}) follows from (\ref{2.175}) as in the proof of Lemma 2.6

Let us now consider the {\it exceptional case}. The new problem is that
 now $m_{\pm}(x, y, 0\pm )\neq 0$. Let us write $\Phi^{(+)}_t$ as
 follows
 \beq
 \Phi^{(+)}_t (x, y)= \Phi^{(1)}_t (x,y) +\Phi^{(2)}_t (x,y),
 \label{2.176}
 \ene
 where
 \beq
 \Phi^{(j)}_t (x,y):=\int_0^{\infty} e^{-ik^2 t}\, \frac{e^{-ik(x-y)}}{2 \pi}
 m^{(j)}(x,y,k)\, dk,\, j=1,2,
 \label{2.177}
 \ene
 with
 \beq
 m^{(1)}(x,y,k):= \Psi (k^2 )\left[q_+ (x,y,k)-q_+ (x,y,0+)\right],
 \label{2.178}
 \ene
 \beq
 m^{(2)}(x,y,k):=\Psi (k^2 )q_+ (x,y,0+).
 \label{2.179}
 \ene
 Then using Theorem 2.5 (b) we prove as in the {\it generic case}\, that
 \beq
 \left|\Phi^{(1)}_t (x,y)\right| \leq \frac{C}{\sqrt{t}},\,x,y \in \ER,\,
 t > 0.
 \label{2.180}
 \ene
 Let $\hat{\Psi}(\lambda ), \lambda \geq 0$, be the cosine transform of
 $\Psi (k^2 )$:
 \beq
 \hat{\Psi}(\lambda ):= \int_0^{\infty} \cos (\lambda k) \Psi(k^2 )dk.
 \label{2.181}
 \ene
 Then integrating by parts we prove that for any $N > 0$ there is a constant
 $C_N$ such that
 \beq
 \left|\hat{\Psi}(\lambda )\right| \leq C_N \, (1+|\lambda |)^{-N}.
 \label{2.182}
 \ene
 Since
 \beq
 \Psi (k^2 )=\frac{2}{\pi} \int_0^{\infty} \cos (\lambda k) \hat{\Psi}(\lambda ) d \lambda ,
 \label{2.183}
 \ene
 we have that
 \beq
 \Phi^{(2)}_t (x,y)=\frac{q_+ (x,y,0+)}{ \pi} \int_0^{\infty} d \lambda
 \hat{\Psi}(\lambda ) \int_0^{\infty} e^{-ik^2 t }e^{-ik(x-y)} \cos(\lambda k)
 \, dk.
 \label{2.184}
 \ene
But

\beq
\left|\int_0^{\infty} e^{-ik^2 t} e^{-ik(x-y)} \cos (\lambda k)\, dk \right|
\leq \frac{C}{\sqrt{t}},\, t > 0.
\label{2.185}
\ene
The estimate (\ref{2.185}) is proven by explicitly evaluating the cosine
transform using the following equations from \cite{8}: 3 in page 7,
1 in page 23, 7 in page 24, 3 in page 63, 1 in page 82 and 3 in page 83.
Then by (\ref{2.182}), (\ref{2.184}) and (\ref{2.185})
\beq
\left|\Phi^{(2)}_t (x,y)\right| \leq \frac{C}{\sqrt{t}},\, x,y \in \ER ,\,
 t > 0.
 \label{2.186}
 \ene
 It follows from (\ref{2.176}), (\ref{2.180}) and (\ref{2.186}) that
 \beq
 \left|\Phi^{(+)}_t (x,y)\right| \leq \frac{C}{\sqrt{t}}, \, x,y \in \ER ,\,
 t > 0.
 \label{2.187}
 \ene

 We prove in the same way that
 \beq
 \left|\Phi^{(-)}_t (x,y)\right|\leq \frac{C}{\sqrt{t}},\, x,y \in \ER ,
 \, t > 0.
 \label{2.188}
 \ene
 Equation (\ref{2.159}) follows from (\ref{2.160}), (\ref{2.161}),
 (\ref{2.187}) and (\ref{2.188}) as in the {\it generic case}.

\noindent {\it Proof of Theorem 1.1:}\, The theorem follows from Corollaries
2.6 and 2.7.

\noindent {\it Proof of Corollary 1.2:}\, Since $H$ is self--adjoint
\beq
\left\|e^{-itH} P_c \right\|_{{\cal B}\left(L^2 \right)} \leq 1.
\label{2.189}
\ene
Then the corollary follows interpolating between (\ref{1.18}) and
(\ref{2.189}) (see the Appendix to \cite{6}).

\noindent{\it Proof of Corollary 1.3:}\, Corollary 1.3 follows from Corollary
1.2 as in the proof of Theorem 4.1 of \cite{10}.

\section{Inverse Scattering}\sss
\noindent {\it Proof of Theorem 1.4:}\, We prove this theorem by verifying
the conditions of the abstract Theorems 1 and 2 of \cite{15} and of
Theorem 16 of \cite{16}. This is done as in Theorem 8 of \cite{15} and Theorem
17 of \cite{16}. We define $X$ and $X_3$ as in the Introduction and
$X_1 :=L^{1+\frac{1}{p}}$. It
 follows from the Sobolev imbedding theorem (see \cite{14}) that
 $X \subset X_3 $, with continuous imbedding. Concerning hyphotesis (V) in page
 113 of \cite{15}: note that since by Sobolev's imbedding theorem
 $W_1 \subset L^{1+p}$; we have that $X_1 \subset W_1$. But as $e^{-itH} \in
 {\cal B}(W_1 )$, it follows by duality that $e^{-itH} \in {\cal B}(W_{-1} )$.
 Then for all $\phi \in X_1$, $e^{-itH}\,\phi \in W_{-1}$ and
  $e^{-itH} \, e^{-isH} \phi =e^{-i(t+s)H} \phi$ for all $t,s \in \ER$.

 To verify hypothesis VII of Theorem 16 of \cite{16}, as in the proof of
 Theorem 8 of \cite{15}, we need the following result. Let $g$ be any
 real--valued $C^2$ function defined on $\ER$ such that $g(0)=0$ and for all
 $u, v \in \ER $:
 \beq
 |g(u)-g(v)|+ |\acute{g}(u)-\acute{g}(v)| \leq C |u-v|,
 \label{3.1}
 \ene
 and
 \beq
 |g(u)|\leq |f(u)|.
 \label{3.3}
 \ene
 For $I$ any interval let us denote by $C(I, X)$ the Banach space of bounded
 and
 continuous functions from $I$ into $X$ with the supremun norm and by
 $B_{\rho} (I,X)$ the ball of center zero and radius $\rho$ in $C(I,X)$.
 Then for any $\phi \in X(\rho /2 )$ and any $s \in \ER$ the equation
 \beq
 u(t)= e^{-itH} \phi +\frac{1}{i} \int_s^t e^{-i(t-\tau )H} P_g (u(\tau ))
 d\tau,
 \label{3.4}
 \ene
 where
 \beq
 P_g (u(\tau )):= g(|u(\tau )|) \frac{u(\tau )}{|u(\tau )|}
 \label{3.5}
 \ene
 has a unique solution $u(t)\in B_{\rho}(\ER ,X)$ and moreover, the $L^2$
 norm and the energy are conserved:
 \beq
 \|u(t)\|_{L^2}=\hbox{constant}
 \label{3.6}
 \ene
 \beq
 E_g := \frac{1}{2} \|\sqrt{H}u(t)\|^2_{L^2}+\int dx G(|u(t)|)=\hbox{constant},
 \label{3.7}
 \ene
 for all $t \in \ER$, where $G$ is the primitive of $g$ such that $G(0)=0$.
 To prove this result we observe that it follows from (\ref{3.1})
 and (\ref{3.3}) that
 \beq
 \| P_g (\phi )-P_g (\psi )\|_X \leq C\left(\|\phi\|_X + \|\psi\|_X \right)
 \|\phi -\psi \|_X ,
 \label{3.8}
 \ene
 for all $\phi ,\psi \in X$. Then by a standard contraction mapping argument
 (\ref{3.4}) as a unique solution on $C([s-\epsilon ,s+\epsilon ],X)$ provided
 that $0 < \epsilon \leq  1/ 3C \rho$ and $ 0 < \epsilon <
 1/2C$. Suppose that (\ref{3.6}) and
 (\ref{3.7}) are true for $t \in [s-\epsilon ,s+\epsilon ]$. Then since
 $|G(\lambda )| \leq C \lambda^2$,
 \beq
 \|u(t)\|_X^2 \leq 2 E_g+ 2 (1+C) \|u(t)\|^2_{L^2} \leq C,\, t \in [s-\epsilon ,
 s+\epsilon ].
 \label{3.9}
 \ene
 Since $\|u(t)\|_X$ remains bounded as $t \rightarrow s \pm \epsilon$ by a
 constant $C$ that depends only on $\|\phi \|_X$ we can extend $u(t)$ into a
 global solution such that (\ref{3.6}), (\ref{3.7}) hold for all $t \in \ER$.
 It remains to prove that (\ref{3.6}), (\ref{3.7}) are true for
 $t \in [s-\epsilon , s+\epsilon ]$. In the constant coefficient case, $V=0$,
 this is accomplished by approximating the local solution in $W_1$ by solutions
 in $W_2$, see \cite{18} and \cite{19} or by regularizing equation (\ref{3.4})
 by taking convolution with a function in Schwartz space, see \cite{20},
 \cite{21} and \cite{22}. This is possible because in the constant
 coefficient case $D(H) = D(\Delta )=W_2$. In our case this is not a convenient
 approach. Since we only assume that $V \in L^1_{\gamma}$ we do not
 have much control over $D(H)$. We only know that $D(H)$ is a dense set in $X$.
 To solve this problem we regularize (\ref{3.4}) multiplying it by an
 appropriate function of $H$. Let us denote
 $r_n (H):= \left(\frac{H}{n}+1 \right)^{-1}, n=1,2,\cdots  $. The regularized
 equation is given by
 \beq
 u_n (t)=e^{-itH} r_n (H) \phi +\frac{1}{i} \int_s^t e^{-i(t -\tau )H}
 r_n (H) P_g \left(r_n (H) u_n (\tau )\right)\, d \tau.
 \label{3.10}
 \ene
 As above we prove that (\ref{3.10}) has a unique solution for
 $t \in [s-\epsilon , s+\epsilon ]$. Note that we can take $\epsilon$
 independent
 on $n$. Moreover, since $H r_n (H) \in {\cal B}(X )$ we have that actually
 $u_n (t) \in C^1 ([s-\epsilon ,s+\epsilon ], X)$. Then
 \beq
 \frac{d}{d t} \|u(t)\|_{L^2}^2 = 2 \Re \left( u_n (t),
 \frac{\partial}{\partial t} u_n (t)\right).
 \label{3.11}
 \ene
 Since $u_n (t)$ is a solution to the equation
 \beq
 i\frac{\partial}{\partial t}u_n (t)=H u_n(t)+
 r_n (H) g\left(|r_n (H) u_n (t)|\right)
 \frac{r_n (H) u_n (t)}{|r_n (H) u_n (t)|}
 \label{3.12}
 \ene
 and since $H$ is self--adjoint, it follows from (\ref{3.11}) that
 \beq
 \frac{d}{d t} \|u_n (t)\|^2_{L^2} = 0.
 \label{3.13}
 \ene
 Furthermore,
 \beq
 \frac{1}{2} \frac{d}{d t} \|\sqrt{H} u_n (t)\|^2_{L^2} = \Re \left( \sqrt{H}
 u_n (t), \sqrt{H} \frac{\partial}{\partial t} u_n (t) \right).
 \label{3.14}
 \ene
 Let us define
 \beq
 Q_n (t):= \int dx \, G\left(|r_n (H) u_n|\right).
 \label{3.15}
 \ene
 Since $|G(\lambda )| \leq C |\lambda |^2$,
 \beq
 |Q_n (t)| \leq C \|u_n (t)\|^2_{L^2}.
 \label{3.16}
 \ene
 Furthermore, since $u_n (t) \in C^1 ([s-\epsilon ,s+\epsilon ], X)$ it follows
 from a simple proof using the fundamental theorem of calculus (see
 the proof of Lemma 3.1 of \cite{20} for a similar argument) that
 \beq
 \frac{d}{d t} Q_n (t)= \Re \left( r_n (H)
 \frac{g(|r_n (H) u_n (t)|)}{|r_n (H) u_n (t)|} r_n (H) u_n (t),
 \frac{\partial}{\partial t} u_n (t) \right).
 \label{3.17}
 \ene
 We define the regularized energy as follows
 \beq
 E_n (t):= \frac{1}{2}\|\sqrt{H} u_n (t) \|^2_{L^2} +Q_n (t).
 \label{3.18}
 \ene
 It follows from (\ref{3.12}), (\ref{3.14}), (\ref{3.17}) and since $H$ is
 self--adjoint that
 \beq
 \frac{d}{d t} E_n (t)=0.
 \label{3.19}
 \ene
 By (\ref{3.13}) and (\ref{3.19}), $\|u_n (t)\|_{L^2}$ and $E_n (t)$ are
 constant for $t \in [s-\epsilon ,s+\epsilon ]$. We prove below that $u_n (t)$
 converges strongly in $X$ to $u(t)$. Since moreover, $r_n (H)$ converges to
 the identity strongly in $X$, equations (\ref{3.6}) and (\ref{3.7}) hold
 for $t \in [s-\epsilon ,s+\epsilon ]$. It only remains to prove
 that
 \beq
 \lim_{n \rightarrow \infty}\|u_n (t) - u(t) \|_X=0.
 \label{3.20}
 \ene
 But by (\ref{3.4}), (\ref{3.8}) and (\ref{3.10})
 $$
 \|u_n (t)-u(t)\|_X \leq \int_s^t d\tau \, \|r_n (H) P_g (r_n (H) u_n )
 -r_n (H) P_g (r_n (H)u)\|_X \,+
 $$
 $$
 \int_s^t d\tau \|r_n (H) P_g  (r_n (H)u)-P_g (u)\|_X \leq
 2 C \epsilon \, \rho \|u_n  -u\|_{C([s-\epsilon ,s+\epsilon ],X)}+
 $$
 \beq
 2 C \epsilon \, \rho \ \int_s^t \|(r_n (H)-1) u(\tau )\|_X \, d\tau .
 \label{3.21}
 \ene
 But since $2 C \epsilon \rho < 2/3$
 \beq
 \|u_n -u\|_{C([s-\epsilon ,s+\epsilon ],X)} \leq  6 C \epsilon \rho
 \int_{s-\epsilon}^{s+\epsilon} \| (r_n (H)-1) u(\tau )\| d\tau \rightarrow 0,
 \label{3.22}
 \ene
 as $n \rightarrow \infty$.
 As in the proof of Theorem 17 of \cite{16} we have to prove that
 $e^{-itH} \in {\cal B}\left( X, L^r \left(\ER , L^{1+p}\right)\right)$.
 Let us denote by ${\cal D}$ the set of points in the
 $(\frac{1}{q}, \frac{1}{r})$ plane, $ 1\leq p,q \leq \infty$, such
 that
 $e^{-itH}\in {\cal B}\left(X,L^r \left( \ER ,L^{q}\right)\right)$.
 We already know that $A:= (\frac{1}{2},0) \in {\cal D}$ because
 $e^{-itH}$ is a unitary  operator on $L^2$. Since $e^{-itH}$ is unitary on
 $X$, we have that $e^{-itH} \in {\cal B}\left( X, L^{\infty}(X)\right)$ and as
 by Sobolev's theorem \cite{14} $X$ is continuously embedded in $L^{\infty}$
 it follows that  $B:= (0,0) \in {\cal D}$. By Corollary 1.3
 $e^{-itH} \in {\cal B}\left( L^2 , L^6 \left(\ER ,L^6 \right)\right)$ and
 then  $C:= (\frac{1}{6},\frac{1}{6}) \in {\cal D}$. Since
 $A, B, C \in {\cal D}$  it follows by interpolation (see \cite{6}) that the
 solid triangle with vertices $A, B, C$ belongs to ${\cal D}$. Let us consider
 the following curve, ${\cal C}$, in the $(\frac{1}{q}, \frac{1}{r})$ plane:
 \beq
 \frac{1}{r}:= \left(\frac{1}{2}+\frac{1}{q}\right)/\left(q-2\right)
 =-\frac{1}{2} +\frac{1}{2-\frac{4}{q}}-\frac{1}{2q},\, 1 \leq q\leq 6.
 \label{3.23}
 \ene
  Note that  ${\cal C}$ goes
 from $B$ to $C$ and that for $0 \leq \frac{1}{q} \leq \frac{1}{6}$ the curve
 ${\cal C}$ is contained in the triangle with vertices $(A, B, C )$. Then
 ${\cal C} \subset {\cal D}$  for $0 \leq \frac{1}{q} \leq \frac{1}{6}$ and
 then taking $q=p-1$, we have that
 $e^{-itH} \in {\cal B}\left( X, L^r \left( L^{p+1}\right)\right)$
 for $5 \leq p \leq \infty$, with $r:= (p-1)/(1-d)$.

 \noindent{ \it Proof of Theorem 1.5 :}\, The proof of Theorem 1.1 of \cite{23}
 applies in our case with no changes.

 \noindent {\it Proof of Corollary 1.6 :} \, By Theorem 1.5 $S$ determines
 uniquely $S_L$. Let us denote
 \beq
 \hat{S_L}:= F S_L F^{\ast}
 \label{3.24}
 \ene
 and let $U$ be the following unitary operator from $L^2$ onto $L^2 (\ER^+ )
 \oplus L^2 (\ER^+ ):$
 \beq
 Uf(k):= \left\{ \begin{array}{c}f_1 (k)\\ f_2 (k)\end{array}\right\},
 \label{3.25}
 \ene

 where $f_1 (k):= f(k), k \geq 0 ,$ and $f_2 (k):= f(-k), k \geq 0.$
 Let us denote
 \beq
 \tilde{S}_L := U \hat{S}_L U^{\ast}.
 \label{3.26}
 \ene
 Pearson proved in Section 9.7 of \cite{24} that for $V$ bounded and with fast
 decay:
 \beq
 \tilde{S}_L  \left\{ \begin{array}{c}f_1 (k)\\ f_2 (k)\end{array}\right\}=
 \left[ \begin{array}{lr}T(k)& R_1 (k)\\R_2 (k) & T(k)\end{array}\right]
  \left[ \begin{array}{c}f_1 (k)\\ f_2 (k)\end{array}\right].
  \label{3.27}
  \ene
  Let us assume that $V \in L^1_{\delta}$ for some $\delta > 1$. Let $V_n \in
  C^{\infty}_0 , n=1,2,\cdots $ be such that
  \beq
  \lim_{n \rightarrow \infty}\left\| V_n -V \right\|_{L^1_{\delta}} =0.
  \label{3.28}
  \ene
  Let us denote by $S_{L,n}, T_n (k)$ and $R_{j,n}(k), j=1,2,$ the
  scattering operator, the transmission coefficient and the reflection
  coefficients corresponding to $V_n$. Then by the proof of Lemma 1 of \cite{1}
  and by equations (\ref{2.60}) to (\ref{2.62})
  \beq
  \lim_{n \rightarrow \infty}T_n (k)= T(k), \,\,
  \lim_{n \rightarrow \infty}R_{j,n}(k)=R_j (k), j=1,2.
  \label{3.29}
  \ene
  Moreover, by the stationary formula for the wave operators (see equation
  (12.7.5) of \cite{13}) and from the results in Chapter 12 of \cite{13}

  \beq
  s-\lim_{n \rightarrow \infty}S_{L,n}=S_L ,
  \label{3.30}
  \ene
  where the limit exists in the strong topology in $L^2$.
  Then by continuity (\ref{3.27}) is true also for
  $V\in L^1_{\delta}, \delta > 1$ and it follows that from $S_L$ we obtain
  the transmission coefficient and the reflection coefficients. But since $V$
  has no bound states one of the reflection coefficients uniquely determines
  $V$ ( see for example \cite{25}, \cite{26}, \cite{1}, \cite{me}
  \cite{28} or \cite{29}).

\noindent {\it Proof of Corollary 1.7:}\,The proof of Corollary 1.3 of
\cite{23}
applies in this case with no changes.

\end{document}